\documentclass[fleqn,usenatbib]{mnras}

\usepackage{newtxtext, newtxmath}
\usepackage{bm}
\usepackage{xcolor}

\usepackage[T1]{fontenc}

\DeclareRobustCommand{\VAN}[3]{#2}
\let\VANthebibliography\thebibliography
\def\thebibliography{\DeclareRobustCommand{\VAN}[3]{##3}\VANthebibliography}

\usepackage{graphicx}	
\usepackage{amsmath}	

\usepackage{multirow}
\usepackage{booktabs}
\usepackage[dvipsnames]{xcolor}
\usepackage{enumitem}
\usepackage{comment}
\newlist{todolist}{itemize}{2}
\setlist[todolist]{label=$\square$}
\usepackage{caption}
\usepackage{subcaption}
\usepackage{arydshln}
\usepackage{soul}
\usepackage{float}
\usepackage{hyperref}
\usepackage[normalem]{ulem}

\newcommand{\secref}[1]{Sec.~\ref{#1}}
\newcommand{\figref}[1]{Fig.~\ref{#1}}
\newcommand{\appref}[1]{Appx.~\ref{#1}}
\newcommand{\tabref}[1]{Tab.~\ref{#1}}
\renewcommand{\eqref}[1]{Eq.~(\ref{#1})}

\title[Strong Lensing and Subhalo Density Profiles]{Using Strong Lensing to Detect Subhalos with Steep Inner Density Profiles}

\author[K. E. Kollmann et al.]{
Kassidy E. Kollmann$^{1}$\thanks{E-mail: kassidy.kollmann@princeton.edu},
James W.\ Nightingale$^{2}$,
Mariangela Lisanti$^{1,3}$,
Andrew Robertson$^{4}$,
Oren Slone$^{5}$
\\
$^{1}$Department of Physics, Princeton University, Princeton, NJ 08544, USA\\
$^{2}$School of Mathematics, Statistics and Physics, Newcastle University, Herschel Building, Newcastle-upon-Tyne, NE1 7RU, UK \\
$^{3}$Center for Computational Astrophysics, Flatiron Institute, New York, NY 10010, USA\\
$^{4}$Carnegie Observatories, 813 Santa Barbara Street, Pasadena, CA 91101, USA\\
$^{5}$Raymond and Beverly Sackler School of Physics and Astronomy, Tel Aviv University, Tel-Aviv 69978, Israel
}

\begin{document}
\label{firstpage}
\pagerange{\pageref{firstpage}--\pageref{lastpage}}
\maketitle
\thispagestyle{empty}

\begin{abstract}
The inner region of a subhalo's density distribution is particularly sensitive to dark matter microphysics, with alternative dark matter models leading to both cored and steeply-rising inner density profiles. This work investigates how the lensing signature and detectability of dark matter subhalos in mock HST-, Euclid-, and JWST-like strong lensing observations depend on the subhalo's radial density profile, especially with regards to the inner power-law slope, $\beta$. We demonstrate that the minimum subhalo mass detectable along the Einstein ring of a system is strongly dependent on $\beta$. In particular, we show that subhalos with $\beta = 2.2$ can be detected down to masses over an order-of-magnitude lower than their Navarro-Frenk-White (NFW) counterparts with $\beta = 1$. Importantly, we find that the detectability of subhalos with steep inner profiles is minimally affected by increasing the complexity of the main lens galaxy's mass model. This is a notable characteristic of these subhalos, as those with NFW or shallower profiles become essentially undetectable when multipole perturbations are added to the lens model. The results of this work highlight how the underlying dark matter physics can significantly impact the expected number of subhalo detections from strong gravitational lensing observations. This is important for testing Cold Dark Matter against alternative models, such as Self-Interacting Dark Matter, that predict a diverse range of subhalo inner density profiles.
\end{abstract}




\section{Introduction}
The Cold Dark Matter~(CDM) model successfully describes the Universe on large cosmological scales~\citep[e.g.,][]{2006Natur.440.1137S, 2020A&A...641A...6P}, and the current challenge is testing it on smaller galactic and sub-galactic scales~\citep{2017ARA&A..55..343B, 2022NatAs...6..897S}. A key prediction of the CDM model is that structure forms hierarchically, with small dark matter halos merging to build larger halos~\citep{1978MNRAS.183..341W}.  In the absence of baryons, the CDM model predicts a universal self-similar density profile for (sub-)halos across all mass scales, which is known as the Navarro-Frenk-White~(NFW) profile~\citep{1996ApJ...462..563N}. This density distribution is a falling double power law that has an inner density slope $\beta=1$, where $\rho(r)\propto r^{-\beta}$ at small radii, and steepens to $\rho(r)\propto r^{-3}$ at larger radii. However, galactic rotation curves show that galaxies exhibit a wide spread of inner density profiles, even for systems of similar mass~\citep{1994ApJ...427L...1F, 1994Natur.370..629M, 2001ApJ...552L..23D, 2004MNRAS.351..903G, 2015MNRAS.452.3650O}. This galaxy diversity poses an important observational test that must be explained by CDM, or any variant thereof.

Many works have investigated if the effects of baryonic feedback can explain this observed diversity in the context of the CDM model~\citep[e.g.,][]{2015MNRAS.452.3650O, 2018MNRAS.473.4392S, 2020MNRAS.495...58S}. For baryonic feedback models with bursty star formation, repetitive and energetic outflows from supernova feedback can redistribute the dark matter in a halo, leading to cored halo density profiles with an inner slope $\beta \sim 0$~\citep{2010Natur.463..203G, 2012MNRAS.421.3464P}. The degree of coring depends on the mass of the subhalo and is most efficient for systems with a stellar-mass-to-halo-mass ratio of $M_*/M_{\rm halo} \sim 5\times10^{-3}$ (e.g., bright dwarfs), but becomes negligible moving to systems with $M_*/M_{\rm halo} \lesssim 5\times10^{-4}$~(e.g., ultrafaint dwarfs)~\citep{2014MNRAS.437..415D, 2016MNRAS.456.3542T, 2020MNRAS.497.2393L}. For more massive systems, the presence of baryons in the halo can lead to density profiles that are steeper than the NFW profile. This is through adiabatic contraction, in which baryonic matter cools and accumulates at the center of the halo, pulling the dark matter inward and increasing its density in the inner region~\citep{1986ApJ...301...27B, 2004ApJ...616...16G}. While baryonic feedback can significantly alter the dark matter distribution in galaxies, these processes are highly mass dependent and may not account for the full spread of density profiles inferred across all halo masses \citep{2015MNRAS.452.3650O, 2018MNRAS.473.4392S, 2020MNRAS.495...58S, 2019MNRAS.487.5272J}. That said, part of the apparent diversity may arise from observational systematics \citep[e.g.,][]{2019MNRAS.482..821O}, while recent work suggests that baryonic feedback within CDM halos can produce greater diversity than previously found \citep{2025arXiv251011800C}.

In addition to feedback effects, the fundamental properties of dark matter at the particle level---such as its mass and interactions---can leave distinctive imprints on galactic scales, affecting both the internal properties of halos and their overall abundance~\citep[e.g.,][]{2012MNRAS.423.3740V, 2013MNRAS.430...81R}. A halo's density distribution is a particularly sensitive probe of dark matter microphysics.  For example, \cite{2019PhRvX...9c1020R} argued that the diversity of inner density profiles observed for galaxies in the Spitzer Photometry and Accurate Rotation Curves \citep[SPARC,][]{2016AJ....152..157L} data set may be explained in the context of the Self-Interacting Dark Matter~(SIDM) model~\citep{2000PhRvL..84.3760S}, where dark matter can interact with itself via a new mediator. These self interactions enable heat to flow within a halo, which results in processes that can significantly alter the dark matter distribution. During the initial core-expansion phase, self interactions transfer heat inward towards the cool central region of the halo, forming an isothermal core \citep{2014PhRvL.113b1302K}. After this stage, the direction of heat flow reverses and, because virialized self-gravitating systems have negative specific heat, the core continues to grow hotter as its radius shrinks, rapidly increasing the central density of the core in a runaway process known as gravothermal core collapse~\citep{1968MNRAS.138..495L, 2002ApJ...568..475B, 2002PhRvL..88j1301B}. The timescale for this process depends on the SIDM model parameters, the halos' structural properties such as mass and concentration~\citep{Essig:2018pzq}, and environmental effects such as tidal stripping~\citep{2020PhRvD.101f3009N}. The halo evolution predicted for SIDM significantly increases the diversity of subhalo density distributions, with profiles ranging from very cored~$(\beta\sim0)$ to very cuspy $(\beta>2)$~\citep{2015MNRAS.453...29E, 2019JCAP...12..010K, 2020PhRvL.124n1102S, 2021MNRAS.505.5327T}.

This paper uses galaxy-galaxy strong gravitational lensing as a probe of dark matter subhalos. In a lensing system, the light from a distant galaxy (the ``source galaxy'') is deflected around a massive foreground galaxy (the ``lens galaxy'') that lies along the line of sight. Depending on the configuration of the system, the source galaxy is multiply imaged and can appear spread out in arcs or a ring. Subhalos within the lens galaxy or along the line of sight can be detected through purely gravitational effects if they cause observable perturbations to the lensed source emission~\citep{2005MNRAS.363.1136K, 2009MNRAS.392..945V, 2009MNRAS.400.1583V}. Strong gravitational lensing is therefore a powerful tool for detecting low-mass subhalos (below $\sim10^8$--$10^9$~M$_\odot$) that remain dark as a result of suppressed star formation due to reionization~\citep[e.g.,][]{2008MNRAS.390..920O, 2016MNRAS.456...85S}. To date, this method has resulted in four low-mass subhalo detections in galaxy-scale strong gravitational lenses: one in SDSS J0946+1006 \citep{2010MNRAS.408.1969V}, two in JVAS B1938+666 \citep{2012Natur.481..341V, 2025NatAs...9.1714P} and one in SDP.81 \citep{Hezaveh2016:2016ApJ...823...37H} (however see \cite{2025A&A...703A.285S} regarding the SDP.81 detection). In the next few years, the number of observed strong lenses is expected to grow by orders of magnitude as a result of surveys such as Euclid \citep{2022A&A...662A.112E, 2025A&A...697A...1E}, which is expected to observe $\sim170,000$ galaxy-galaxy strong lenses~\citep{2015ApJ...811...20C, 2025arXiv250315324E}, with $\sim 2500$ expected to yield substructure detections~\citep{ORiordan2023}. Furthermore, recent data from the James Webb Space Telescope~(JWST) \citep{2006SSRv..123..485G, 2023PASP..135f8001G} has led to the detection of over 100 strong lens candidates with sources that extend to redshifts $z>6$, higher than any other lens survey~\citep{2025arXiv250308777N}. 

When using strong lensing observations to detect dark matter subhalos, a common approach for modeling the subhalo is to assume a CDM profile, such as the NFW profile with a mass-concentration relation, or a tidally truncated pseudo-Jaffe (PJ) profile~\citep{2001ApJ...558..657M}. In both cases, the subhalo's mass and two-dimensional position in the lens plane are the only free parameters of the subhalo model. This approach simplifies the fitting procedure by limiting the number of free parameters; however, doing so also limits what can be learned about the subhalo and the underlying distribution of dark matter. In recent years, this approach has been extended to additionally include the subhalo's concentration as a free parameter of the NFW model. It has been shown that not accounting for scatter in the mass-concentration relation can bias the inferred subhalo mass \citep{2021MNRAS.507.1202M} and the expected number of subhalo detections for a given dark matter model~\citep{Amorisco2022:2022MNRAS.510.2464A}. \cite{2021MNRAS.507.1202M} also demonstrated that mass and concentration can be simultaneously constrained for massive or very compact subhalos when both properties are fit for in the model. Furthermore, multiple studies have explored using various subhalo profiles to model the detected subhalos in order to better constrain their density profiles. Such works have found that three of the detected subhalos are well fit by steep density profiles with inner slopes that are close to or steeper than isothermal \citep{2021MNRAS.507.1662M, 2025A&A...699A.222D, Minor2025, 2025MNRAS.540..247E, 2025MNRAS.543..540T, 2025arXiv251011006Y, 2025NatAs...9.1714P}.\footnote{However, \cite{2025ApJ...991L..53H} re-analyzed the subhalo detection in  SDSS J0946+1006 and found that including a luminous satellite galaxy component in the lens model can result in a shallower subhalo density profile than previously inferred.} These subhalos appear to be outliers of the CDM model and have motivated significant interest in using strong gravitational lensing to probe the nature of dark matter \citep[e.g.,][]{2021MNRAS.507.1202M, 2021MNRAS.507.1662M, 2021MNRAS.507.2432G, Amorisco2022:2022MNRAS.510.2464A, 2022MNRAS.516..336S, 2023ApJ...958L..39N, 2025A&A...699A.222D, Minor2025, 2025JCAP...08..048H, 2025MNRAS.540..247E, 2025ApJ...978...38D, 2025MNRAS.543..540T, 2025arXiv250820179O, 2025arXiv251001491K, 2025arXiv251011006Y, 2025NatAs...9.1714P}.

In this work, we investigate how the lensing signature and detectability of non-CDM subhalos differs from their CDM counterparts. We specifically test how the detectability of subhalos in a galaxy-galaxy strong lensing system depends on the subhalo's inner density slope, a characteristic known to be highly impacted by complex dark matter physics. We simulate mock strong lensing observations that contain a subhalo perturber with a density distribution that ranges from strongly cored to steeply rising. We model the data using a flexible subhalo profile with three free parameters (mass, concentration, and inner density slope) and investigate how robustly the true subhalo properties can be recovered. We explore how these results depend on the position of the subhalo in the lens galaxy, the properties of the observing instrument, and the complexity of the main lens galaxy's mass model. The SIDM model provides an excellent case study for this paper, as it naturally leads to both cored and cuspy subhalo density profiles. This work is therefore phrased in the context of a CDM versus SIDM subhalo detectability study. However, we note that SIDM is not the only model of relevance.  In particular, scenarios where the dark matter dissipates energy in its interactions can lead to cuspy density profiles, while scenarios where the dark matter is an ultralight boson can lead to cores \citep[e.g.,][]{2019PhRvL.123n1301M, 2021MNRAS.506.4421S, 2022SciA....8J3618C, 2023MNRAS.524..288O, 2025ApJ...982..175R}.

This paper is organized as follows. \secref{sec:mock_data} introduces the mock observations simulated for this work, which vary the properties of the subhalo perturber and of the observing instrument used to simulate the data. We create mock observations that resemble HST-, Euclid-, and JWST-like data. \secref{sec:fitting} describes the lens-modeling procedure performed on all observations and how we search for the presence of subhalos. \secref{sec:results} presents the results of our primary analysis. In \secref{sec:pixelized}, we re-analyze a subset of the mock observations using more advanced modeling techniques to corroborate the results of our main analysis. Finally, we discuss and summarize our results in \secref{sec:discussion_conclusions}. Supplementary figures are included in \appref{sec:appendix}. This work uses the open-source strong lensing software \texttt{PyAutoLens}\footnote{https://github.com/Jammy2211/PyAutoLens}~\citep{Nightingale2018:2018MNRAS.478.4738N, Nightingale2021b:2021JOSS....6.2825N} for all strong lensing analyses and assumes a~\cite{2016A&A...594A..13P} cosmology.

\section{Mock Data} 
\label{sec:mock_data}
For this work, we create mock observations of a galaxy-galaxy strong lensing system with a single subhalo perturber within the main lens galaxy. There are two key steps to creating mock strong lensing observations with \texttt{PyAutoLens}. The first is to specify parameterizations for all light and mass components of the system. These parameterizations are used in ray-tracing calculations to compute how the system would appear to an observer after light from the source galaxy has been deflected by the foreground mass distribution. The second step is to specify instrumental properties that control how the simulated lens system is observed, and the quality of the mock data. This includes setting the exposure time of the observation, the background sky level, the pixel scale of the data, and the properties of the point-spread function~(PSF) that the image is convolved with.  

The deflection of light from a background source due to a foreground mass distribution is described by the lens equation:
\begin{equation} \label{eqn:lens_eqn}
    \boldsymbol{\beta}=\boldsymbol{\theta}-\boldsymbol{\alpha}(\boldsymbol{\theta}) \, ,
\end{equation}
where $\boldsymbol{\beta}=(\beta_1, \beta_2)$ are the angular coordinates in the plane of the source,\footnote{It is standard notation to represent the source plane coordinates in the lens equation as $\boldsymbol{\beta}=(\beta_1, \beta_2)$.  However, we emphasize that this is completely different from the $\beta$ used to represent the subhalo's inner density slope.} $\boldsymbol{\theta}=(\theta_1, \theta_2)$ are the angular coordinates in the plane of the lens, and $\boldsymbol{\alpha}(\boldsymbol{\theta})=(\alpha_1, \alpha_2)$ are the deflection angles that describe how the source light is deflected by the foreground mass distribution. In this work, we follow $\texttt{PyAutoLens}$ conventions and use $(x, y)$ to represent all angular coordinates. Elliptical profiles are written in terms of the elliptical coordinate, $\xi$:
\begin{equation}
\xi  =\sqrt{x^2 + y^2/q^2} \, ,
\end{equation}
and are characterized by the ellipticity components: 
\begin{equation} \label{ell_comps}
    \epsilon_{\scriptscriptstyle{1}} = \frac{1-q}{1+q}\sin2\phi  \hspace{0.1in} \text{and} \hspace{0.1in} \epsilon_{\scriptscriptstyle{2}} = \frac{1-q}{1+q}\cos2\phi \, ,
\end{equation}
where $q$ is the (minor-to-major) axis ratio and $\phi$ is the angle of the major axis relative to the positive $x$-axis. In what follows, \secref{sec:macro} discusses details of the light and mass profiles used to simulate the source galaxy and main lens galaxy, which are collectively part of the ``macro" system.  \secref{sec:sub_perturber} then discusses the density profile used to simulate the subhalo perturber and how the subhalo's properties are varied across the observations. Finally, \secref{sec:suites} discusses how the mock data is organized into five suites, which differ from each other either in regards to the instrument properties or the location of the subhalo within the lens system.

\subsection{Macro System} \label{sec:macro}
The simulated system consists of a source galaxy at a redshift of 1.0 and a lens galaxy at a redshift of 0.2 that lies directly along the line of sight between the observer and the source galaxy. The lens system used in this study is in a standard cross quad lensing configuration, in which the background source galaxy is lensed into four multiple images. The properties of the macro system are consistent with fits to true observed lens systems, such as the Sloan Lens ACS~(SLACS) lenses~\citep{2006ApJ...638..703B, 2008ApJ...682..964B, 2010ApJ...724..511A, 2024MNRAS.52710480N}, but are not chosen to emulate any particular system. This generality is intentional, as this paper is focused on testing how variations to the \textit{subhalo's} properties impact its detectability. A natural follow-up to this work would be to perform the analysis on a larger sample of simulated lenses, systematically varying properties of the macro system and/or matching to observed lens systems.   

We model all light components using an elliptical cored-S\'ersic distribution~\citep{Graham2003:2003AJ....125.2951G, Trujillo2004:2004AJ....127.1917T} which combines an outer S\'ersic  profile~\citep{1963BAAA....6...41S} with an inner power-law function. The intensity of this profile is given by
\begin{equation} \label{eqn:Sersic}
    I(\xi)=I'\left(1+\left(\frac{R_b}{\xi}\right)^{\alpha'}\right)^{\gamma'/
    \alpha'}\exp\left[-b\left(\frac{\xi+R_b}{R_e}\right)^{1/n}\right] \, ,
\end{equation}
where
\begin{equation}\label{eqn:Iprime}
    I'=I_b\,2^{-\gamma'/\alpha'}\exp\left[b\, 2^{1/\alpha' n}(R_b/R_e)^{1/n}\right] \, .
\end{equation}
The cored-S\'ersic profile is characterized by six parameters: $R_{b}$ is the break radius where one profile transitions to the other, $\alpha'$ is the sharpness of the profile transition, $\gamma'$ is the slope of the inner power-law function, $n$ is the S\'ersic index, $I_b$ is the intensity at the break radius, and $R_e$ is the half-light radius of the profile. The parameter $b$ is a function of the other profile parameters and is used to set $R_e$ as the half-light radius. The intensity of a standard S\'ersic profile diverges to infinity at its center, which can cause numerical issues for lens modeling. We therefore use a \textit{cored}-S\'ersic profile for all light components with core parameters fixed to: $\alpha' = 3.0$, $\gamma' = 0.25$, and $R_{b}$ equal to half the pixel scale of the observation. These parameters are chosen to ensure that the light profiles are cored enough to prevent any pixels from having problematically high light intensities, while also being consistent with observed lens systems~\citep{Trujillo2004:2004AJ....127.1917T}. Throughout the fitting procedure, the core parameters of the profile are held fixed. 

The cored-S\'ersic light profile is a steep profile with large surface brightness gradients near the center. Consequently, the intensity of source light can vary significantly within a single image-plane pixel. To accurately compute how much light from the source galaxy contributes to each image-plane pixel, we use \emph{adaptive oversampling}, which divides each pixel into a finer sub-grid. The position of each sub-pixel is ray traced to the source plane and the intensity of all sub-pixels is averaged to compute the intensity of the overall pixel. This process is repeated, dividing the pixel into an increasingly higher resolution sub-grid until a threshold accuracy is achieved where further dividing the pixel no longer has any significant impact on the average intensity of that pixel. This process is also referred to as ``supersampling'' and has been shown to be important for robust lens modeling~\citep{Minor2025}.

The source galaxy's light is simulated using one elliptical cored-S\'ersic profile. Two cored-S\'ersic profiles are used to simulate the main lens galaxy's light, as observed lenses are typically massive elliptical galaxies, which are more realistically fit with multiple light components~\citep{2019MNRAS.489.2049N, 2024MNRAS.532.2441H}. We refer to the two light components of the main lens galaxy as the bulge and disk. \tabref{tab:macro_params} lists the cored-S\'ersic profile parameters used to simulate the source and main lens galaxy's light, with superscripts $\scriptscriptstyle{\rm S}$, $\scriptscriptstyle{\rm bulge}$, and $\scriptscriptstyle{\rm disk}$ used to indicate the corresponding component. The intensity of the light profiles is scaled by a factor $\delta$ to achieve the desired signal-to-noise ratio~(SNR) of the data, which is discussed in \secref{sec:suites}. 

\begin{table*}
 \renewcommand{\arraystretch}{1.2} 
 \begin{tabular*}{2\columnwidth}{@{}l@{\hspace*{10pt}}l@{\hspace*{7pt}}l@{\hspace*{10pt}}l@{\hspace*{7pt}}l@{\hspace*{10pt}}l@{}}
 
  \toprule
  
\textbf{Component} & \textbf{Parameter} & \textbf{Description} & \textbf{True Value} & \textbf{Prior Type} & \textbf{Prior Range}  \\
  
  \midrule
  
  \multirow{5}{*}{Source Galaxy Light} & $(x, y)^{\scriptscriptstyle{\rm S}}$ & Center position of profile $\left[\arcsec \right]$ & $(0,\ 0)$ & Gaussian & $\mu=0,\ \sigma=0.3$\\
  
  & $(\epsilon_{\scriptscriptstyle{{\rm 1}}}, \epsilon_{\scriptscriptstyle{{\rm 2}}})^{\scriptscriptstyle{{\rm S}}}$ & Ellipticity & $(0.10,\ -0.06)$ & Gaussian & $\mu=0,\ \sigma=0.3$\\
  
  & $I_b^{\scriptscriptstyle{{\rm S}}}$ & Intensity at break radius $\left[e^{-}{\rm/pix/s}\right]$ & $3\delta$ & -- & --\\
  
  & $R_e^{\scriptscriptstyle{{\rm S}}}$ & Effective radius $\left[\arcsec \right]$ & $0.1$ & Uniform & $[0,\ 30]$\\
  
  (1 elliptical cored S\'ersic) & $n^{\scriptscriptstyle{{\rm S}}}$ & S\'ersic index & $2$ & Uniform & $[0.8,\ 5]$\\
  
  \midrule
  
  \multirow{8}{*}{Lens Galaxy Light} & \multirow{2}{*}{$(x, y)^{\scriptscriptstyle{\rm bulge,disk}}$} & \multirow{2}{*}{Center position of profile $\left[\arcsec \right]$} & Bulge: $(0,\ 0)$ & \multirow{2}{*}{Gaussian} & \multirow{2}{*}{$\mu=0,\ \sigma=0.3$}\\

  & & & Disk: $(0,\ 0)$ & & \\
  
  & \multirow{2}{*}{$(\epsilon_{\scriptscriptstyle{{\rm 1}}}, \epsilon_{\scriptscriptstyle{{\rm 2}}})^{\scriptscriptstyle{{\rm bulge,disk}}}$} & \multirow{2}{*}{Ellipticity} & Bulge: $(0,\ -0.05)$ & \multirow{2}{*}{Gaussian} & \multirow{2}{*}{$\mu=0,\ \sigma=0.3$}\\
  
  & & & Disk: $(0.02,\ -0.05)$ & & \\
  
  & \multirow{2}{*}{$I_b^{\scriptscriptstyle{{\rm bulge,disk}}}$} & \multirow{2}{*}{Intensity at break radius $\left[e^{-}{\rm /pix/s}\right]$} & Bulge: $2.8\delta$ & \multirow{2}{*}{--} & \multirow{2}{*}{--}\\

  & & & Disk: $2.8\delta$ & & \\
  
  & \multirow{2}{*}{$R_e^{\scriptscriptstyle{{\rm bulge,disk}}}$} & \multirow{2}{*}{Effective radius $\left[\arcsec\right]$} & Bulge: $0.7$ & \multirow{2}{*}{Uniform} & \multirow{2}{*}{$[0,\ 30]$}\\

  & & & Disk: $4.5$ & & \\
  
  & \multirow{2}{*}{$n^{\scriptscriptstyle{{\rm bulge,disk}}}$} & \multirow{2}{*}{S\'ersic Index} & Bulge: $3$ & \multirow{2}{*}{Uniform} & \multirow{2}{*}{$[0.8,\ 5]$}\\
  
  (2 elliptical cored S\'ersics) & & & Disk: $1.2$ & & \\
  
  \midrule

  \multirow{4}{*}{Lens Galaxy Mass} & $(x, y)^{\scriptscriptstyle{{\rm L}}}$ & Center position of profile $\left[\arcsec \right]$ & $(0,\ 0)$ & Gaussian & $\mu=0,\ \sigma=0.3$\\
  & $(\epsilon_{\scriptscriptstyle{{\rm 1}}}, \epsilon_{\scriptscriptstyle{\rm 2}})^{\scriptscriptstyle{{\rm L}}}$ & Ellipticity & $(0.11,\ 0)$ & Gaussian & $\mu=0,\ \sigma=0.3$\\
  & $\theta_{\scriptscriptstyle{E}}^{\scriptscriptstyle{{\rm L}}}$ & Einstein radius $\left[\arcsec \right]$ & $1.4$ & Uniform & $[0,\ 8]$\\
  (elliptical power-law) & $\eta^{\scriptscriptstyle{{\rm L}}}$ & Power-law slope & $2.1$ & Uniform & $[1.5,\ 3]$\\
  \midrule
  External Shear & $(\gamma_1, \gamma_2)^{\scriptscriptstyle{{\rm ext}}}$ & Shear components & $(0.05,\ 0.05)$ & Uniform & $[-0.2,\ 0.2]$\\
  \midrule
  \multirow{3}{*}{Multipole Perturbations} & \multirow{3}{*}{$(\epsilon_{\scriptscriptstyle{{\rm 1}}}, \epsilon_{\scriptscriptstyle{{\rm 2}}})^{\scriptscriptstyle{{\rm m}}}$} & $m=1$ components & -- & Uniform & $[-0.2,\ 0.2]$\\
  & & $m=3$ components & -- & Uniform & $[-0.1,\ 0.1]$\\
  & & $m=4$ components  & -- & Uniform & $[-0.1,\ 0.1]$\\
  \midrule
  
  \multirow{9}{*}{Subhalo Mass} & \multirow{2}{*}{$(x, y)^{\scriptscriptstyle{{\rm sub}}}$} & \multirow{2}{*}{Center position of profile $\left[\arcsec \right]$} & On-Ring: $(0.3,\ 1.3)$ & \multirow{2}{*}{Gaussian} & \multirow{2}{*}{$\mu=\rm truth,\ \sigma = 1$} \\
  & & & Offset: $(0.3,\ 2.3)$ & & \\ 
  & \multirow{2}{*}{$\log_{\rm 10}(M_{\rm 200}$/M$_\odot)$} & \multirow{2}{*}{Subhalo mass} & \multirow{2}{*}{Varied: $[7,\ 10]$} & \multirow{2}{*}{Uniform} & \multirow{2}{*}{$[6,\ 11]$} \\
  & & & & & \\
  & \multirow{2}{*}{$c_{\rm gNFW}$} & \multirow{2}{*}{Concentration} & \multirow{2}{*}{Varied: \cite{2016MNRAS.460.1214L}} & \multirow{2}{*}{LogUniform} & \multirow{2}{*}{$[1,\ 100]$}\\  
  & & & & & \\
  & \multirow{3}{*}{$\beta$} & \multirow{3}{*}{Inner slope} & Cored: $0.2$ & \multirow{3}{*}{Uniform} & \multirow{3}{*}{$[0,\ 2.5]$}\\
  & & & NFW: $1$ & & \\
  (generalized NFW) & & & Steep: $2.2$ & & \\
\bottomrule
 \end{tabular*}
  \caption{This table lists the light and mass components of the lens system used in this work. The parameters associated with each component are described in the second and third columns. Superscripts are used on the parameters to indicate which component they correspond to, as discussed in the text. For brevity, we omit superscripts on the subhalo's mass, concentration, and inner slope. The fourth column lists the value of each parameter used to simulate the mock observations. The relative intensity of the source and lens light profiles is held fixed for all suites, but we vary the intensity of all light profiles by a factor $\delta$ to achieve the desired signal-to-noise ratio of the data, as discussed in \secref{sec:suites}. The fifth and sixth columns list the prior type and prior range used for each component in the lens-modeling process. For all light profiles used in this work, we hold fixed three parameters related to the core of the cored-S\'ersic profile: $R_{b} = \frac{1}{2}$ of the pixel scale of the observation, $\alpha' = 3.0$, and $\gamma' = 0.25$. The lens galaxy's light is subtracted from the data before modeling in our primary analysis, but is included in \secref{sec:pixelized}. No prior information is included for the intensity of the light profiles as these are solved for via linear inversion and are not free parameters of the \texttt{Nautilus} search. We use a Gaussian prior for the subhalo's position centered at the subhalo's true location in the lens plane. Redshifts are not fit for in the modeling process. Multipole perturbations are not used to simulate the lens system, but are included in part of the modeling procedure (see \secref{sec:smooth}).}
   \label{tab:macro_params}
\end{table*}

The total mass distribution of the main lens galaxy is simulated using an elliptical power-law~(EPL) density profile~\citep{EPL:2015A&A...580A..79T, 2020MNRAS.496.3424O}. The convergence of this profile (i.e., the surface mass density scaled by the critical lensing surface mass density) is given by
\begin{equation}\label{eq:lens_mass}
    \kappa(\xi) = \frac{(3-\eta)}{1+q}\left(\frac{\theta_{\scriptscriptstyle{E}}}{\xi}\right)^{\eta-1} \, ,
\end{equation}
where $q$ is the axis ratio of the profile, $\eta$ is the three-dimensional power-law slope, and $\theta_{\scriptscriptstyle{E}}$ is the Einstein radius~\citep{1998ApJ...502..531B}. We also include an external shear component that contributes additional deflections of the source light caused by galaxies along the line of sight or by unmodelled complexities of the main lens galaxy's mass distribution~\citep{2024MNRAS.531.3684E}. In \texttt{PyAutoLens}, the external shear is parameterized by two components $(\gamma_1, \gamma_2)^{\rm ext}$, which relate to the magnitude of the shear, $\gamma^{\rm ext}$, and the position angle, $\phi^{\rm ext}$:
\begin{equation} \label{eqn:shear}
    \gamma^{\rm ext} = \sqrt{(\gamma_1^{\rm ext})^2 + (\gamma_2^{\rm ext})^2} \quad\quad \text{and} \quad\quad \phi^{\rm ext}=\frac{1}{2}\arctan\left(\frac{\gamma_2^{\rm ext}}{\gamma_1^{\rm ext}}\right) \, .  
\end{equation}
The properties of the main lens galaxy's mass distribution and the external shear are listed in \tabref{tab:macro_params} with superscripts $\scriptscriptstyle{\rm L}$ and $\scriptscriptstyle{\rm ext}$ used respectively.

\begin{figure*}
    \centering
    \includegraphics[scale=1]{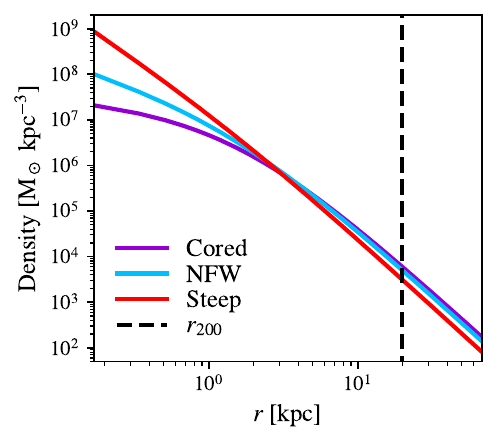}
    \includegraphics[scale=1]{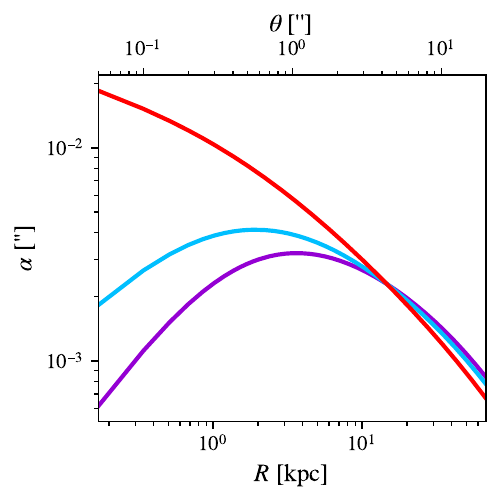}
    \caption{\emph{Left:} Dark matter density profiles as a function of three-dimensional  radius, $r$, for a subhalo of mass $M_{\rm 200} = 10^9$~M$_\odot$ and a concentration ($c_{\rm gNFW}$) of 13.5, which is set based on the \protect\cite{2016MNRAS.460.1214L} mass-concentration relation.  The results are shown for the Cored~($\beta=0.2$), NFW~($\beta=1$), and Steep~($\beta=2.2$) subhalo profiles in purple, blue, and red, respectively. The black dashed line marks the $r_{200}$ radius for these subhalos. \emph{Right:} Subhalo deflection-angle strength as a function of projected distance from the subhalo, $R$, for the same three subhalos as on the left. On the top axis, we use the angular diameter distance to the lens plane to convert the physical distance to angular distance. For small $R$, the subhalo's deflection strength strongly depends on its inner slope.}
    \label{fig:def_curve_dens}
\end{figure*}

\subsection{Subhalo Perturber} \label{sec:sub_perturber}
We include a single subhalo perturber in the main lens galaxy to study how its properties, specifically its inner density slope, affect the subhalo's lensing signature and detectability. The subhalo's dark matter density distribution is modeled using a spherical generalized NFW~(gNFW) profile, which is given by
\begin{equation} \label{eqn:gNFW}
    \rho_{\scriptscriptstyle{\rm gNFW}}(r) = \frac{\rho_{\scriptscriptstyle{0}}}{\left(\frac{r}{r_s}\right)^\beta \left(1+\frac{r}{r_s}\right)^{3-\beta}} \, ,
\end{equation}
where $\rho_{\scriptscriptstyle{0}}$ is the density normalization, $r_s$ is the scale radius, and $\beta$ is the inner density slope. This profile reduces to the standard NFW profile in the case where $\beta=1$. The gNFW profile is characterized by three parameters: the subhalo's mass, concentration~$(c_{\scriptscriptstyle{{\rm gNFW}}} = r_{200}/{r_s})$, and inner density slope.  Here, $r_{200}$ is the radius within which the gNFW profile has an average density equal to 200 times the critical density of the Universe, where $\rho_{\rm crit}=3H^2/8\pi G$ with $H$ being the Hubble parameter and $G$ Newton's constant. The subhalo mass definition used in this work is $M_{\rm 200}$, which is the mass the subhalo would enclose within a radius $r_{200}$ if it were a field halo and not tidally stripped. Real subhalos experience tidal forces as they orbit their host halo, which can strip mass from their outer regions. A subhalo's strong lensing signal is primarily sensitive to the mass enclosed within its inner region, at radii much smaller than those impacted by tidal stripping. For example, \cite{2021MNRAS.507.1202M} found no apparent bias in the subhalo's inferred mass and concentration when using an untruncated NFW $(\beta=1.0)$ profile to fit a tidally stripped NFW subhalo perturber in a strong lens system. In this work, we simulate and fit all subhalos using an untruncated gNFW profile. A natural extension to this work would investigate if the inferred subhalo parameters are biased when an untruncated gNFW density profile is used to fit tidally stripped subhalos with varying inner density slopes. Additionally, one could perform the analysis presented here using a truncated gNFW density profile to simulate and fit subhalos, with the subhalo's tidal radius as an additional free parameter in the modeling.

\begin{figure*}
    \centering
    \includegraphics[width=\textwidth]{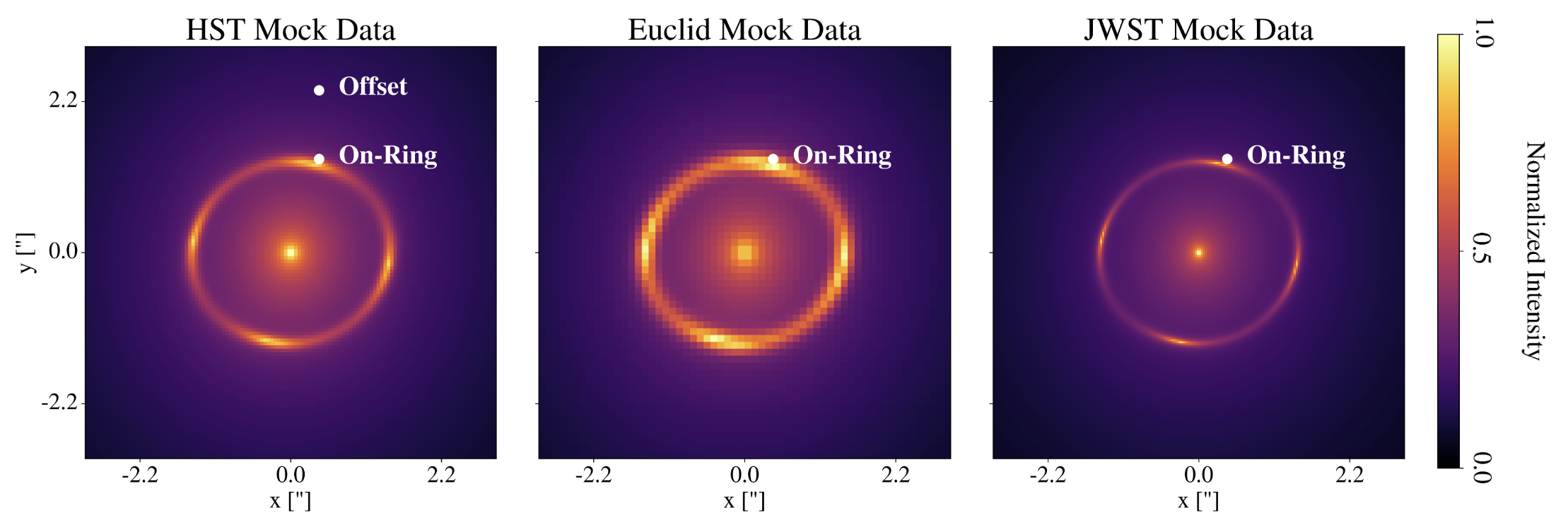}
    \caption{Mock HST- (left), Euclid- (center), and JWST-like (right) observations of the macro system described in \secref{sec:macro}. Subhalos in the HST-like data are positioned at a projected location in the lens plane of either (0.3\arcsec, 1.3\arcsec) or (0.3\arcsec, 2.3\arcsec), labeled as On-Ring and Offset, respectively. Subhalos in the Euclid- and JWST-like data are positioned only at the On-Ring position. For each image, the color bar represents the intensity of the data normalized by the maximum intensity of each image. The images shown here are noiseless (see \secref{sec:fitting} for further discussion).}
    \label{fig:system_images}
\end{figure*}

When varying the subhalo parameters, we use twenty different values for its mass, which are sampled logarithmically in the range $M_{\rm 200} = 10^{7.5}$--$10^{10}$~M$_\odot$. Although subhalos at the higher end of this mass range would likely host a satellite galaxy, we only simulate and model the subhalo's dark matter component. The subhalo's concentration is set to the median value from the \cite{2016MNRAS.460.1214L} mass-concentration relation based on the value of the subhalo's mass. We also vary the subhalo's inner density slope, $\beta$, using three different values: $\beta = 0.2,\ 1.0,\ 2.2$. The three different slopes are rough proxies for the various stages of SIDM halo evolution:\footnote{Note that we use the CDM mass-concentration relation even for the Cored and Steep subhalos. This is a common assumption made in the literature when modeling SIDM subhalos; it is a reasonable starting point if the self interactions only affect the central regions of the subhalos.}

\vspace{0.1in}
    \noindent$\bullet$ $\beta=0.2$ represents subhalos in the core-expansion phase,\vspace{0.05in}
    
    \noindent$\bullet$ $\beta=1.0$ represents subhalos with an NFW density profile, as predicted by CDM (or SIDM models with a sufficiently small cross section),\vspace{0.05in}
    
    \noindent$\bullet$ and $\beta=2.2$ represents subhalos that have undergone gravothermal core collapse.\vspace{0.1in}
    
\noindent Subhalos with these inner slopes will be referred to as ``Cored'', ``NFW'', and ``Steep'', respectively. The left panel of \figref{fig:def_curve_dens} shows the gNFW density profiles for three $10^{9}$~M$_\odot$ subhalos, spanning the three different inner slopes. The black dashed line marks the $r_{200}$ radius for these subhalos.

For each of these density profiles, the right panel of \figref{fig:def_curve_dens} shows the corresponding deflection-angle strength as a function of the projected radial distance, $R$, from the subhalo's center, which is computed as follows. For a spherically symmetric mass distribution, the surface mass density can be obtained by integrating the three-dimensional density profile, $\rho(r)$, along the line of sight:
\begin{equation} \label{surfdens}
    \Sigma(R)=2\int_0^\infty\rho(r)dz = 2\int_R^\infty \frac{\rho(r)rdr}{\sqrt{r^2 - R^2}} \, ,
\end{equation}
where $r=\sqrt{R^2+z^2}$. The mass enclosed within a projected distance $R$ from the profile center is then given by
\begin{equation} \label{projmass}
    M(R)=2\pi \int_0^R\Sigma(R') R' dR' \,.
\end{equation}
The angular amount by which this mass distribution deflects the background source's light is then
\begin{equation} \label{alpha_hat_eqn}
    \hat{\alpha}(R) = \frac{4G M(R)}{c^2 R}\, .
\end{equation}
The reduced deflection angle is defined as
\begin{equation} \label{alpha_eqn}
    \alpha(R) \equiv \frac{D_{LS}}{D_S}\hat{\alpha}(R),
\end{equation}
where $D_S$ is the angular diameter distance to the source plane and $D_{LS}$ is the angular diameter distance between the source and lens planes. The reduced deflection angle, $\alpha$, is what appears in \eqref{eqn:lens_eqn}. Throughout this work, any discussion of deflection strength or perturbation strength refers to the reduced deflection angle.

The right plot of \figref{fig:def_curve_dens} is useful for understanding how much each subhalo deflects light from the source galaxy, based on the impact parameter of the light ray relative to the subhalo’s center. This distance is shown in physical units in the lens plane on the bottom horizontal axis and is converted to angular distance (using the angular diameter distance to the lens plane) on the top horizontal axis. The shape of the Steep subhalo's deflection curve is noticeably different from that of the Cored and NFW subhalos. This is a consequence of how the deflection strength depends on the logarithmic slope of the subhalo's density profile and subsequently on the amount of mass enclosed within radii below the scale radius. To gain intuition for this, consider a power-law density profile $\rho(r)\propto r^{-k}$ so that $\Sigma(R)\propto R^{1-k}$ $\Rightarrow$ $M(R)\propto R^{3-k}$, and $\Rightarrow$ $\alpha(R)\propto R^{2-k}$. The gNFW density profile is $\propto r^{-\beta}$ in the limit of small radii and $\propto r^{-3}$ in the large-radius limit. For the NFW and Cored subhalos, therefore, the deflection strength increases with distance from the subhalo for small $R$ and decreases with distance for large radii. For the Steep subhalo with $\beta = 2.2$, the density slope is always $>2$, and so the deflection strength always decreases with increasing $R$ and has its maximal value at $R\rightarrow 0$ (i.e., at the closest position to the subhalo). As can be seen in \figref{fig:def_curve_dens}, the three deflection curves start to converge to similar values as $R$ increases. Moving towards smaller $R$, however, causes the relative deflection between the Cored and Steep subhalos to diverge. Varying the subhalo's inner density slope most significantly impacts how the subhalo perturbs source emission in the immediate vicinity of the subhalo. This fact will be used in the following section to justify the  choice of where to position the subhalo within the main lens galaxy when simulating the mock data.

\begin{table*}
 \renewcommand{\arraystretch}{1.2} 
 \begin{tabular*}{2\columnwidth}{@{}l@{\hspace*{15pt}}l@{\hspace*{15pt}}l@{\hspace*{15pt}}l@{\hspace*{15pt}}l@{\hspace*{15pt}}l@{}}
  \toprule
  \textbf{Property} & \textbf{HST On-Ring} & \textbf{HST Offset} & \textbf{HST On-Ring} & \textbf{Euclid On-Ring} & \textbf{JWST On-Ring} \\
  & & & \textbf{(Higher SNR)} & &\\
  \midrule
  Subhalo Position $\left[\arcsec \right]$ & (0.3, 1.3) & (0.3, 2.3) & (0.3, 1.3) & (0.3, 1.3) & (0.3, 1.3) \\
  
  Pixel Scale $\left[\arcsec{\rm /pix} \right]$ & 0.05 & 0.05 & 0.05 & 0.1 & 0.031 \\

  PSF FWHM $\left[\arcsec \right]$ & 0.1 & 0.1 & 0.1 & 0.17 & 0.035 \\
  
  Background Sky Level $\left[e^{-}{\rm /s/pix}\right]$ & 0.1 & 0.1 & 0.1 & 1.2 & 0.04 \\

  Exposure time [s] & 2400 & 2400 & 9600 & 2260 & 1800\\

  SNR$_{\rm max}$ & $\sim$30 & $\sim$30 & $\sim$60 & $\sim$30 & $\sim$70 \\

  \hdashline

  Lens Model Fit & Fiducial \& Multipole & Fiducial \& Multipole & Fiducial & Fiducial & Fiducial \\
  \bottomrule
   \end{tabular*}
   \caption{Summary of the mock data suites used in this work. The variations between the different suites are related to the subhalo position or observation parameters. The last row summarizes how the main lens galaxy's mass distribution is modeled for that suite in the primary analysis (see \secref{sec:smooth}).}
 \label{tab:suite_properties}
\end{table*}

\subsection{Mock Data Suites} \label{sec:suites}
The mock observations used in this work are organized into five suites. The observations within a particular suite are identical to one another except for the mass, concentration, and inner slope of the subhalo perturber. We vary over the twenty subhalo masses (which simultaneously varies the concentration) and the three values of $\beta$ discussed in \secref{sec:sub_perturber}, for a total of 60 mock observations per suite with a different subhalo perturber in each. The five suites are different from one another in regards to either the location of the subhalo within the main lens galaxy, the observing instrument used to simulate the data, and/or the maximum SNR, SNR$_{\rm max}$, of the observation. In this work, SNR$_{\rm max}$ corresponds to the SNR of the brightest pixel in the lensed arc when light from the main lens galaxy is subtracted from the data. 

Firstly, we create our benchmark suite which consists of mock observations designed to resemble observations from the Hubble Space Telescope/Advanced Camera for Surveys~(HST/ACS). The data in this suite is created with a pixel scale of 0.05\arcsec, a Gaussian PSF with a full-width-at-half-maximum~(FWHM) of 0.10\arcsec, an exposure time of 2400~s, and a background sky level of 0.1~$e^{-}$/s/pixel~\citep{2007ApJS..172..196K}. We vary the intensity of the source and lens galaxy's light to achieve a SNR$_{\rm max}\sim30$, which is representative of many lensed sources observed in HST surveys such as SLACS~\citep{2008ApJ...682..964B}. The left panel of \figref{fig:system_images} shows a mock HST-like observation of the macro system. For this benchmark suite, the subhalo is located at an angular position of (0.3\arcsec, 1.3\arcsec), which projects the subhalo to lie along the top of the Einstein ring. This ``On-Ring'' position ensures that there is source emission in the immediate vicinity of the subhalo. Since varying the subhalo's inner slope primarily affects the subhalo's deflection strength at small $R$, this ``HST On-Ring'' suite will be used to investigate the lensing signature of different subhalo profiles. Next, we create the ``HST Offset'' suite, which is identical to the HST On-Ring suite in terms of instrument/observation properties, and differs only in regards to the subhalo's position. In this suite, the subhalo is located at an angular position of (0.3\arcsec, 2.3\arcsec), which projects the subhalo to lie $\sim1$\arcsec~from the Einstein ring. This is marked and labeled as the ``Offset'' position in the left panel of \figref{fig:system_images}. The motivation for this suite is to investigate how the sensitivity to the subhalo's inner density slope decreases as the subhalo is moved further from the Einstein ring. 

The remaining three suites all place the subhalo back at the On-Ring position, but vary properties of the observing instrument/observation. The ``HST On-Ring~(Higher SNR)'' suite still emulates HST-like observations, but has an exposure time 4$\times$ longer than the HST On-Ring suite to achieve a SNR$_{\rm max}$ that is doubled to $\sim60$. We use this suite to investigate how improving the SNR of the data enhances the sensitivity to lower-mass subhalos. The remaining two suites differ from the HST On-Ring suite in terms of the observation type, with one suite emulating observations from Euclid~\citep{2011arXiv1110.3193L, 2016SPIE.9904E..0QC} and the other emulating JWST~\citep{jwst_nircam_docs} observations. The ``Euclid On-Ring'' mock observations are created with a pixel scale of 0.1\arcsec, a Gaussian PSF FWHM of 0.17\arcsec, an exposure time of 2260~s, and a background sky level of 1.2~$e^{-}$/s/pixel. We vary the intensity of the light profiles to achieve an SNR$_{\rm max}\sim30$, which is representative of lenses in the Euclid Q1 sample~\citep{2025arXiv250315324E, 2025arXiv250315325E}. The ``JWST On-Ring'' mock observations are created with a pixel scale of 0.031\arcsec, a Gaussian PSF FWHM of 0.035\arcsec, an exposure time of 1800~s, and a  background sky level of 0.04~$e^{-}$/s/pixel, with the light profile intensities varied so that SNR$_{\rm max}\sim70$. This SNR$_{\rm max}$ can be realistically obtained for dedicated strong lensing follow-up observations with JWST. The center (right) panels of \figref{fig:system_images} show mock Euclid-like (JWST-like) observations of the macro system. The five mock data suites used in this work are summarized in Tab.~\ref{tab:suite_properties}.

\section{Lens Modeling and Subhalo Search}
\label{sec:fitting}
To model a lens system, parameterizations are assumed for all light components of the system and for the mass distribution responsible for the lensing. A sampler for Bayesian posterior and evidence estimation is then used to explore the full parameter space of the model. \texttt{PyAutoLens} uses a technique called non-linear search chaining, which utilizes the probabilistic programming language \texttt{PyAutoFit}\footnote{https://github.com/rhayes777/PyAutoFit}~\citep{2021JOSS....6.2550N} and the samplers \texttt{Nautilus}~\citep{2023MNRAS.525.3181L} and \texttt{Dynesty}~\citep{2020MNRAS.493.3132S} to fit a sequence of lens models of increasing complexity to the mock data. The results of fitting each model are then used to initialize the priors for the subsequent, more complex lens model that is fit. The Source, Light, and Mass~(SLaM) Pipelines of \texttt{PyAutoLens} can be used to fit a smooth lens model (i.e., a model without a subhalo) to the data. The full SLaM pipeline sequence is listed below with a brief description of each pipeline's main purpose (see \cite{2022RAA....22b5014C, 2022MNRAS.517.3275E, He:2023MNRAS.518..220H,2024MNRAS.52710480N} for a detailed description of these pipelines):\vspace{0.1in}

    \noindent (i) Parametric Source Pipeline: uses parametric profiles to initialize a model for all light and mass components.\vspace{0.05in}
    
    \noindent (ii) Pixelized Source Pipeline: reconstructs the source galaxy's light on a pixelized grid (important for properly modeling complex source morphologies).\vspace{0.05in}
    
    \noindent (iii) Lens Light Pipeline: re-models the lens galaxy's light with the lens mass fixed based on the results of the previous pipeline.\vspace{0.05in}
    
    \noindent (iv) Lens Mass Pipeline: re-models the lens galaxy's mass using a more complex mass model.\vspace{0.1in}

The pixelized source reconstruction is important for analyzing real strong lensing observations, as source galaxies can have complex, asymmetric morphologies that cannot be properly modeled by a parametric light profile. However, since the source galaxy in our data is simulated with a cored-S\'ersic profile, we use a simplified version of the SLaM pipelines that models the source's light parametrically throughout the full procedure. Additionally, since the light from the main lens galaxy is visibly distinct from the source galaxy's light in our data, we subtract it prior to performing the fitting procedure. The main analysis procedure therefore uses the Parametric Source Pipeline and the Lens Mass Pipeline (\secref{sec:smooth}), followed by a Subhalo Pipeline, which is used to search for a subhalo perturber in each observation (\secref{sec:sub_search}). This approach is used to model all mock observations in this work to build intuition for how subhalo detectability depends on the subhalo's inner density slope using data that varies the subhalo's position and the properties of the observation.  The modeling simplifications made here make the results easier to interpret and also significantly reduce the computational time needed to scan over all variations. In \secref{sec:pixelized}, we re-analyze a subset of the mock observations using the full SLaM Pipeline procedure (with the Subhalo Pipeline) to ensure that our results hold when using the more complicated modeling procedure necessary for analyzing real data.

The Bayesian evidence, $\varepsilon$, is the probability of the data, $\bm{D}$, given the model, $\bm{M}(\bm{\Theta})$, marginalized over the prior distribution of the parameters, $\bm{\Theta}$:
\begin{equation}
    \varepsilon \equiv p \left( \bm{D} \, | \, \bm{M} \right) = \int \mathcal{L}\left( \bm{D} \, | \, \bm{\Theta}, \bm{M} \right) p \left(\bm{\Theta} \, | \, \bm{M} \right) d {\bm \Theta} \,.
\end{equation}
Here, $\mathcal{L}\left( \bm{D} \, | \, \bm{\Theta}, \bm{M} \right)$ is the likelihood and $p \left(\bm{\Theta} \, | \, \bm{M} \right)$ is the prior distribution.  The (natural) log-likelihood function\footnote{Jupyter notebooks that provide a visual step-by-step guide of the {\tt PyAutoLens} likelihood function used in this work can be accessed at \url{https://github.com/Jammy2211/autolens_likelihood_function}.} used to characterize how well a parametric model with a specific set of parameters $\bm{\Theta}$ fits the data is given by:
\begin{equation}\label{eqn:likelihood}
    \ln \mathcal{L}\left(\bm{D} \, | \, \bm{\Theta}, \bm{M} \right) = -\frac{1}{2} \sum_{i} \left( \frac{D_i - M_i \left(\bm{\Theta}\right)}{N_i} \right)^2  + \text{const.}\, ,
\end{equation}
where $D_i$ and $M_i(\bm{\Theta})$ are the flux in the $i^{\rm th}$ image pixel of the data image and predicted model image, respectively. $N_i$ corresponds to the noise on the total photon count in the $i^{\rm th}$ image pixel, which for this work, is assumed to be uncorrelated with the noise of other pixels.

When a mock observation is simulated, Poisson noise is typically added to the image by random draw from a Poisson distribution based on the total photon count in each pixel from the lens signal and the background sky. The corresponding noise map for each observation is then computed as the square root of the observed random realization of the data. $D_i$ and $N_i$ in \eqref{eqn:likelihood} are therefore both affected by the specific random realization. Consequently, the random realization of the data can impact the best-fit lens model inferred for a given mock observation and can introduce scatter in the posterior probability distributions inferred for the free parameters of the model. When working with mock data, this scatter can be accounted for by working with multiple versions of the data with different random noise realizations and averaging the results of all versions. Another option, as used by~\cite{Amorisco2022:2022MNRAS.510.2464A}, is to work with noise-free mock observations, in which a random realization of Poisson noise is not added to the data image. The corresponding noise map used when fitting a lens model to the data is then equal to the square root of the noise-free data image. In this work, we employ this noise-free data approach, meaning no randomness exists in $D_i$ and $N_i$ based on a specific realization of the data. This approach allows our posterior probability distributions to be more easily interpretable, as they should peak at the input parameter values.

\subsection{Smooth Model Fit} \label{sec:smooth}
The first step of the fitting procedure is the Parametric Source Pipeline, which fits a smooth model to the data using parametric profiles for all light and mass components. Since the lens light is subtracted prior to modeling, the only light component to model is that of the source galaxy. This is modeled using the cored-S\'ersic profile discussed in \secref{sec:macro}, which is characterized by the seven free parameters listed in \tabref{tab:macro_params}. The intensity of the source galaxy's light profile is solved for via a linear inversion using the method described by \cite{2024MNRAS.532.2441H}, which builds on the work of \cite{2003ApJ...590..673W} and uses the fast non-negative least-square (fnnls) algorithm\footnote{The fnnls code we are using is modified from \url{https://github.com/jvendrow/fnnls}.} \citep{Bro1997} to ensure a positive solution. This approach fits the intensity separately from the other model parameters, reducing the dimensionality of the parameter space \texttt{Nautilus} searches through. The other free parameters of this search are those of the main lens galaxy's mass model and the two free parameters for the external shear. For this initial search, the lens galaxy's mass distribution is fit using an EPL profile with $\eta^{\scriptscriptstyle{L}} = 2$ (equivalent to an elliptical isothermal mass profile) with its central position fixed to (0\arcsec, 0\arcsec). In total, \texttt{Nautilus} fits 11 free parameters for the Parametric Source Pipeline: 
\begin{equation} \label{eq:theta_source}
    \bm{\Theta}^{\scriptscriptstyle{\rm Parametric\ Source\ Pipeline}} = \{x^{\scriptscriptstyle{S}}, y^{\scriptscriptstyle{S}}, \epsilon_{\scriptscriptstyle{\rm 1}}^{\scriptscriptstyle{S}}, \epsilon_{\scriptscriptstyle{\rm 2}}^{\scriptscriptstyle{S}}, R_e^{\scriptscriptstyle{S}}, n^{\scriptscriptstyle{S}}, \epsilon_{\scriptscriptstyle{\rm 1}}^{\scriptscriptstyle{L}}, \epsilon_{\scriptscriptstyle{\rm 2}}^{\scriptscriptstyle{L}},  \theta_{\scriptscriptstyle{E}}^{\scriptscriptstyle{L}},\gamma_1^{\scriptscriptstyle{\rm ext}}, \gamma_2^{\scriptscriptstyle{\rm ext}}  \}.   
\end{equation}
The priors for all model parameters are listed in \tabref{tab:macro_params} and are chosen to encapsulate measured properties from observed lens systems, such as SLACS~\citep{2008ApJ...682..964B}.

In \secref{sec:macro}, an oversampling grid was used to ensure numerical convergence when computing the cored-S\'ersic light intensity in each pixel of the mock data. We again use an oversampling grid, now to ensure an accurate calculation of the cored-S\'ersic light intensity in each image-plane pixel of the \textit{model image}. This oversampling grid is created after running the Parametric Source Pipeline and is applied to the model image calculation in the subsequent Lens Mass Pipeline and Subhalo Pipeline. The sub-grid resolution of each pixel for this oversampling grid is determined based on how far from the source galaxy's center each image-plane pixel maps to in the source plane. Pixels that map closer to the source galaxy's center are divided into a higher-resolution sub-grid than those that map further away, in order to ensure that the model accurately evaluates the light intensity for very bright regions in the image plane. This distance is calculated by ray tracing each image-plane pixel back to the source plane using the maximum log-likelihood model inferred from the Parametric Source Pipeline and the inferred location of the source galaxy's center. 

Next, we run the Lens Mass Pipeline, which fits a more complex model for the lens galaxy's mass while simultaneously refining the cored-S\'ersic model for the source galaxy's light. We use two different mass models from this step onward. Firstly, for all mock data suites, we use the ``Fiducial'' model, where the lens galaxy's mass is modeled using an EPL mass profile. The free parameters of this model are the same as those from the Parametric Source Pipeline, but with the addition of the power-law slope, $\eta^{\scriptscriptstyle{L}}$, and the center position of the lens galaxy's mass profile. Therefore, there are a total of 14 free parameters in the Fiducial model of the Lens Mass Pipeline:
\begin{eqnarray} \label{eq:theta_mass_fid}
    \bm{\Theta}^{\scriptscriptstyle{\rm Lens\ Mass\ Pipeline,\ Fiducial}} &=& \{\bm{\Theta}^{\scriptscriptstyle{\rm Parametric\ Source\ Pipeline}}, x^{\scriptscriptstyle{L}}, y^{\scriptscriptstyle{L}}, \eta^{\scriptscriptstyle{L}}  \} \, . \nonumber \\
\end{eqnarray}
The Lens Mass Pipeline uses the results of the Parametric Source Pipeline to refine the priors for the source galaxy and lens galaxy parameters that were previously fit. The priors for the three new EPL parameters are given in \tabref{tab:macro_params}.

While the main lens galaxy in this work is simulated as an EPL mass profile, observations of real elliptical galaxies indicate that their isophotes exhibit deviations from pure ellipticity~\citep{1989A&A...217...35B, 2006MNRAS.370.1339H}, with strong lens mass models also showing this~\citep{2024A&A...688A.110S, Barone2024, 2025MNRAS.539..704L}. To account for this complexity, one can model the lens' mass distribution using an extension of the EPL profile that includes multipole perturbations~\citep{2013ApJ...765..134C}. In this case, the functional form of the convergence is given in polar coordinates by
\begin{equation}
    \kappa(R, \phi) = \frac{1}{2}\left(\frac{\theta_{\scriptscriptstyle{E}}}{R}\right)^{\eta-1} k_m \cos(m(\phi - \phi_m)) \, ,
\end{equation}
where $k_{\scriptscriptstyle{m}}$ is the multipole strength, $\phi_{\scriptscriptstyle{m}}$ is the multipole position angle, and $m$ is the multipole order. Previous works have demonstrated that failing to model the true complexity of the lens galaxy's mass distribution can lead to false-positive subhalo detections, where the unmodeled lens mass complexity is mistaken for a subhalo perturbation~\citep[e.g.,][]{He:2023MNRAS.518..220H, ORiordan:2024MNRAS.528.1757O}. Including multipole perturbations in the main lens galaxy's mass model is therefore an important component when modeling real lens systems. However, this degeneracy between the macro model and subhalo component goes both ways and including multipoles in the lens model can result in reduced detectability of true subhalos~ \citep{ORiordan:2024MNRAS.528.1757O, 2025arXiv250902660O}.

To investigate how our detectability results are affected by using a more complex lens model, we re-analyze all observations in the HST On-Ring and HST Offset suites with 1st-, 3rd-, and 4th-order multipoles ($m=1, 3,$ and $4$) included in the main lens galaxy's mass model. This is referred to as the ``Multipole'' model and is used to re-analyze these observations from the Lens Mass Pipeline onward. In \texttt{PyAutoLens}, each $m^{\rm th}$ multipole perturbation is characterized by two parameters ($\epsilon_{\scriptscriptstyle{\rm 1}}$, $\epsilon_{\scriptscriptstyle{\rm 2}})^{\scriptscriptstyle{m}}$, related to $k_{\scriptscriptstyle{m}}$ and $\phi_{\scriptscriptstyle{m}}$ through
\begin{equation} \label{mult_comps}
    k_m=\sqrt{ {(\epsilon_{\scriptscriptstyle{{\rm 1}}}^{\scriptscriptstyle{m}}})^2+{(\epsilon_{\scriptscriptstyle{\rm 2}}^{\scriptscriptstyle{m}}})^2}, \ \ \ \ \ \ \ \ \ \ \ \ \phi_{\scriptscriptstyle{m}} = \frac{1}{m}\arctan\left(\frac{\epsilon_{\scriptscriptstyle{\rm 2}}^{\scriptscriptstyle{m}}}{\epsilon_{\scriptscriptstyle{\rm 1}}^{\scriptscriptstyle{m}}}\right) \, .  
\end{equation}
The Multipole model therefore has a total of 20 free parameters:
\begin{eqnarray} \label{eq:theta_mass_mult}
    \bm{\Theta}^{\scriptscriptstyle{\rm Lens\ Mass\ Pipeline,\ Multipole}} &=& \{\bm{\Theta}^{\scriptscriptstyle{\rm Parametric\ Source\ Pipeline}}, x^{\scriptscriptstyle{L}}, y^{\scriptscriptstyle{L}}, \eta^{\scriptscriptstyle{L}}, \epsilon_{\scriptscriptstyle{\rm 1}}^{\scriptscriptstyle{m=1}}, \nonumber \\
    && \quad \quad 
    \epsilon_{\scriptscriptstyle{\rm 2}}^{\scriptscriptstyle{m=1}}, \epsilon_{\scriptscriptstyle{\rm 1}}^{\scriptscriptstyle{m=3}}, \epsilon_{\scriptscriptstyle{\rm 2}}^{\scriptscriptstyle{m=3}}, \epsilon_{\scriptscriptstyle{\rm 1}}^{\scriptscriptstyle{m=4}}, \epsilon_{\scriptscriptstyle{\rm 2}}^{\scriptscriptstyle{m=4}}  \} \, . \nonumber \\
\end{eqnarray}
The priors for the multipole parameters are given in \tabref{tab:macro_params}. These priors encapsulate the magnitude of multipoles inferred from elliptical isophote fitting of massive elliptical galaxies~\citep{2006MNRAS.370.1339H, 2025MNRAS.540.3281A}. Note that we do not include the monopole~($m=0$) contribution as it is degenerate with a rescaling of the Einstein radius, nor do we include the quadrupole~($m=2$) as it is degenerate with a rescaling of the lens axis ratio and position angle.

\subsection{Subhalo Search} \label{sec:sub_search}
After fitting the smooth model to the data using the Parametric Source Pipeline and the Lens Mass Pipeline, we then run the Subhalo Pipeline. The model fit in the Subhalo Pipeline consists of all 14 or 20 free parameters from the Lens Mass Pipeline (depending on whether the Fiducial or Multipole mass model was used) plus an additional set of parameters for the subhalo component. The subhalo perturber is modeled using the gNFW profile introduced in Section \ref{sec:sub_perturber}, which adds five additional free parameters to the model, corresponding to the subhalo's mass, concentration, inner density slope, and two-dimensional position in the lens plane. There are a total of 19 or 25 free parameters in the Subhalo Pipeline:
\begin{eqnarray} \label{eq:theta_sub}
    \bm{\Theta}^{\scriptscriptstyle{\rm Subhalo\ Pipeline}} = \{\bm{\Theta}^{\scriptscriptstyle{\rm Lens\ Mass\ Pipeline}}, x^{\scriptscriptstyle{\rm sub}}, y^{\scriptscriptstyle{\rm sub}}, M_{\scriptscriptstyle{\rm 200}}, c_{\scriptscriptstyle{\rm gNFW}}, \beta\} \, . 
\end{eqnarray}
The Subhalo Pipeline uses the results of the Lens Mass Pipeline to initialize priors for the smooth model parameters that were previously fit. The priors for the new subhalo parameters added to the model are given in \tabref{tab:macro_params} and are chosen to span a wide range of subhalo mass, concentration, and inner density slopes. When analyzing real data, the Subhalo Pipeline first performs a grid search that divides the image plane into smaller square segments and searches for the presence of a subhalo in each segment. The subhalo position from the maximum log-likelihood fit of the grid search is then used to initialize Gaussian priors on the subhalo's position. Performing the subhalo grid search is not necessary for our study, so we instead use a Gaussian prior for the subhalo's $(x,y)$ position centered at the subhalo's true location, with a large standard deviation of 1\arcsec. 

To determine if the data prefers a model with or without a subhalo, we compute the following statistic:\footnote{This is the natural logarithm of the Bayes factor, $B = \varepsilon_{\rm sub}/\varepsilon_{\rm smooth}$.}
\begin{equation} \label{eqn:delta_ln_evidence}
\Delta\ln\varepsilon = \ln \varepsilon_{\rm sub} - \ln\varepsilon_{\rm smooth}  \, ,  
\end{equation}
where $\varepsilon_{\rm smooth}$ is the Bayesian evidence of the smooth model and $\varepsilon_{\rm sub}$ is the Bayesian evidence of the model with a subhalo component. Before computing $\Delta\ln\varepsilon$, we re-run the smooth model fit once more using the updated priors based on the Lens Mass Pipeline results. This is done to ensure that the same prior width is used for all free parameters that are present in both models being compared in \eqref{eqn:delta_ln_evidence}. A positive value of $\Delta\ln\varepsilon$ means that the model with a subhalo is preferred over the smooth model.  We use an increase of $\Delta\ln\varepsilon = 50$ as the threshold for a strong subhalo detection, as has been adopted in other studies \citep[e.g.,][]{2014MNRAS.442.2017V, Ritondale2019a, Despali2022, 2025A&A...699A.222D}. This threshold corresponds to a decisive detection for a subhalo by the Jeffreys' Scale.

\section{Results}
\label{sec:results}

 \begin{figure}
    \centering
    \includegraphics[]{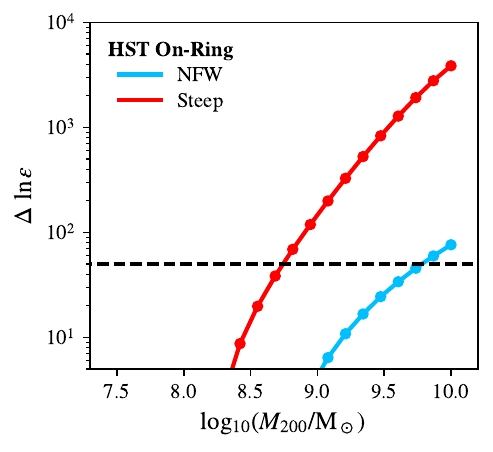}
    \caption{Difference between the total marginalized Bayesian evidence of the lens model with and without a subhalo component as a function of the true mass of the subhalo in the data. Points are only shown for data sets from the HST On-Ring suite with $\Delta\ln\varepsilon> 5$, which results in no Cored subhalos being shown. Blue (red) points correspond to data sets with an NFW (Steep) subhalo. The black dashed line marks the $\Delta\ln\varepsilon=50$ threshold used for detectability. Steep subhalos become detectable at a mass of $5.4\times10^8$~M$_\odot$, while NFW subhalos remain undetectable until a mass of $6.0\times10^9$~M$_\odot$, over an order of magnitude above the minimum detectable Steep subhalo mass.}
    \label{fig:detectability}
\end{figure}

\begin{figure*}
    \centering
    \includegraphics[width=0.49\textwidth]{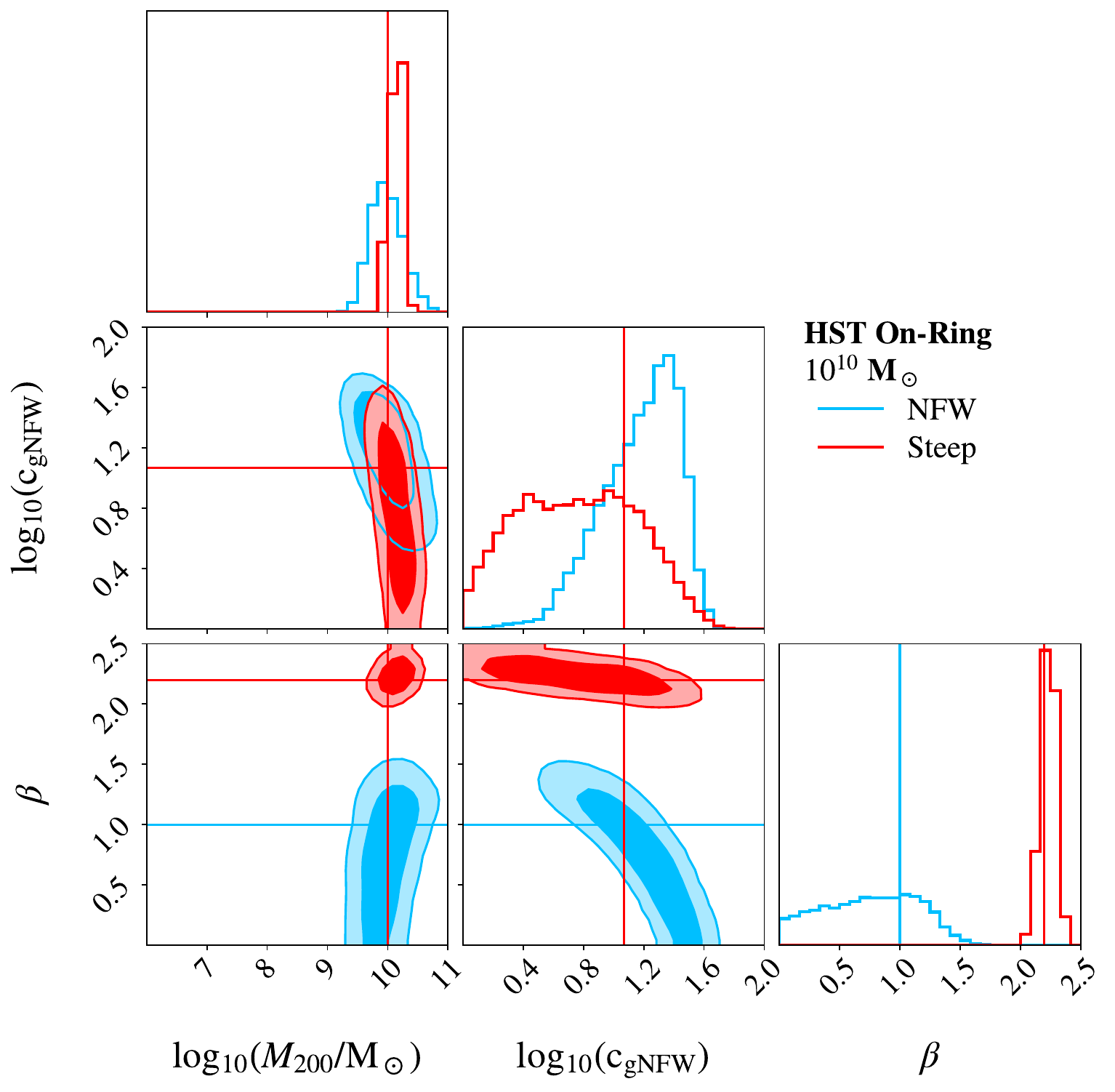}
    \includegraphics[width=0.49\textwidth]{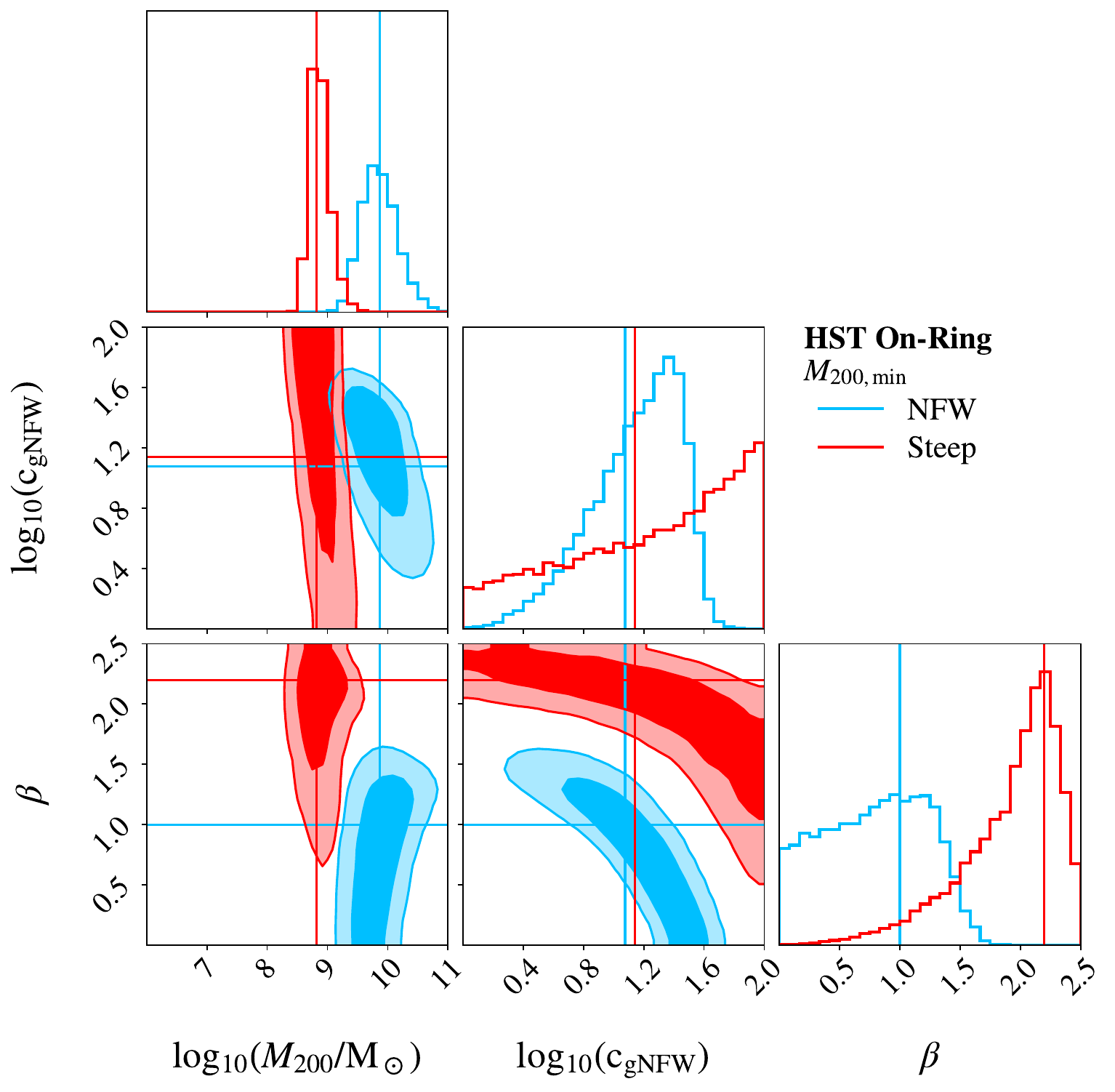}
    \caption{Posterior probabilities for the subhalo parameters inferred for the HST On-Ring Steep~(red) and NFW~(blue) subhalos. The left plot is for the $10^{10}$ M$_\odot$ Steep and NFW subhalos. The right plot is for the subhalo in the HST On-Ring suite with a mass closest to the minimum value detectable for each density slope. The true value of each parameter is marked by the vertical line of the corresponding color. Since both the Steep and NFW subhalos in the left corner plot correspond to the same subhalo mass and concentration, there is only one vertical truth line shown for each. The 2D contours correspond to the 68\% and 95\% confidence regions. In both cases, the subhalo mass and inner slope are well-recovered, with the uncertainties decreasing at higher mass. The posterior distributions for concentration are not constrained as well, but still consistent with the true values.}
    \label{fig:HST_OnRing_cornerplots}
\end{figure*}

We perform the lens-modeling and subhalo search procedure discussed in \secref{sec:fitting} on all mock observations in the five suites. For each suite, we investigate how the cored versus cuspy nature of the subhalo's inner density slope impacts subhalo detectability. For detectable subhalos, we investigate how accurately their true properties can be inferred from the data. We first present the analysis results for the HST On-Ring suite in \secref{sec:key_results}. After building intuition using this benchmark example, we investigate how variations to the subhalo position~(\secref{sec:offset}), observational parameters~(\secref{sec:quality_results}), and complexity of the lens model~(\secref{sec:mult_results}) impact the results.

\begin{figure*}
    \centering
    \includegraphics[width=\linewidth]{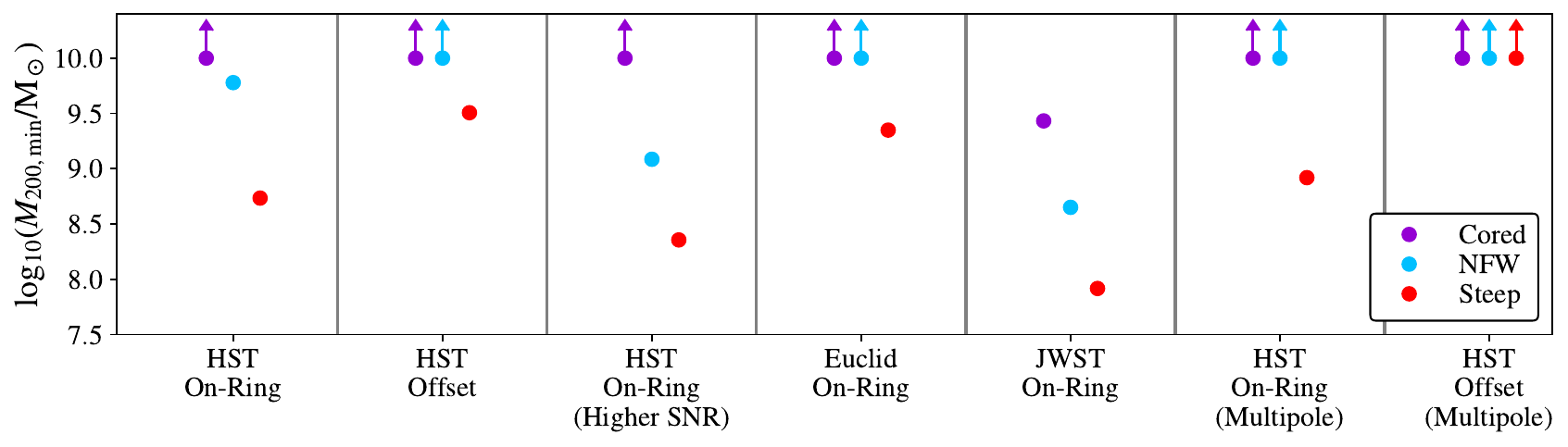}
    \caption{Minimum Cored~(purple), NFW~(blue), and Steep~(red) subhalo mass detectable for all variations explored in this work. An upper arrow is used to indicate when subhalos remain below the detectability threshold even for the maximum subhalo mass used ($10^{10}$ M$_\odot$). Steep subhalos are significantly more detectable in all variations except for the HST Offset (Multipole) case. Remarkably, the minimum-mass detectable for Steep subhalos is barely affected by adding multipoles to the lens model.}
    \label{fig:M_lowest}
\end{figure*}

\subsection{Benchmark Results: HST On-Ring}
\label{sec:key_results}

\figref{fig:detectability} shows the difference in the total marginalized Bayesian evidence between the lens model fit with a subhalo component compared to that without, as a function of the subhalo perturber's true mass for the HST On-Ring suite. The NFW and Steep subhalos are indicated by blue and red curves, respectively. Points are shown only for data sets where $\Delta \ln \varepsilon \geq5$, and the detectability threshold ($\Delta \ln \varepsilon=50$) is marked by the horizontal dashed line. For all masses, Steep subhalos have the highest $\Delta \ln \varepsilon$, followed by the NFW and then the Cored subhalos.\footnote{This is consistent with \cite{2022MNRAS.516..336S}, which found that subhalos with steeper effective density slopes are more detectable.} (All data sets with Cored subhalos have $\Delta\ln\varepsilon < 5$ and are not visible on this plot.) In particular, the Steep subhalos become detectable at a mass of $5.4\times10^8$~M$_\odot$, while the NFW subhalos do so at $6.0\times10^9$~M$_\odot$, over an order of magnitude higher in mass. These results highlight the impact that $\beta$, and hence the underlying dark matter physics, has on subhalo detectability. This suggests substantial differences in the number of expected subhalo detections for CDM compared to SIDM models with different cross sections.

We next investigate how accurately the true properties of detectable subhalos are recovered. \figref{fig:HST_OnRing_cornerplots} shows two corner plots of the posterior distributions for $M_{\rm 200}, c_{\scriptscriptstyle{\rm gNFW}}$, and $\beta$ inferred from the HST On-Ring suite, with true values marked by the vertical lines. Results are provided only for the NFW and Steep subhalos since the Cored subhalos are undetectable. The left panel corresponds to the NFW and Steep subhalos with a true mass of $10^{10}$~M$_\odot$. As this is the largest mass considered for the subhalo perturber, the parameter recovery for these subhalos should be the best-case scenario for this suite. Indeed, the masses for the Steep and NFW cases are well recovered, with $\log_{10}(M_{\rm 200, Steep}/$M$_\odot)=10.2^{+0.1}_{-0.1}$ and $\log_{10}(M_{\rm 200, NFW}/$M$_\odot)=10.0^{+0.3}_{-0.3}$, respectively.\footnote{In this work, all uncertainties are quoted at a 68\% confidence.} The Steep subhalo's inner slope is inferred to be $\beta_{\rm steep}=2.24^{+0.06}_{-0.07}$, which confidently rules out a cored or NFW  profile. In contrast, for the NFW subhalo, the inner slope is constrained to be $\beta_{\rm NFW}=0.77^{+0.40}_{-0.47}$, which is broad enough that there is a non-negligible chance of mistaking this for a more cored subhalo. Specifically, this posterior distribution predicts a $67\%$ probability for the subhalo to have an inner slope $< 1$. However, the probability of mistaking the NFW subhalo for Steep is very unlikely, with only a $2\%$ chance of inferring an inner slope $>2$. 

The $\beta$–$\log_{10}(\rm c_{\rm gNFW})$ panel shows that the inferred concentration and inner slope are correlated in the posterior, reflecting a degeneracy in how these parameters trade off to reproduce a similar lensing signal. Of these two correlated parameters, the inner slope is more robustly inferred. For the $10^{10}$~M$_\odot$ subhalos, the recovered concentrations are  $c_{\scriptscriptstyle{\rm gNFW, Steep}}=5.8^{+9.5}_{-3.7}$ and $c_{\scriptscriptstyle{\rm gNFW, NFW}}=16.4^{+10.2}_{-8.4}$; both are consistent with the true value of 11.7, but with notable spread. The correlation between slope and concentration can be understood as follows. Consider a gNFW subhalo of a given mass with $\beta<3$ and a density $\rho_0$ at some radius $r_0$, where $r_0<r_{\rm s}$. Increasing $\beta$ or $c_{\rm gNFW}$ causes the density at $r_0$ (and the subhalo's enclosed mass within $r_0$ and corresponding deflection strength) to increase. Therefore, increasing either parameter while keeping $\rho_0$, or the enclosed mass within $r_0$, fixed forces the other parameter to decrease. This results in the banana-shaped posterior (blue contour) in the slope-concentration subpanel. For smaller values of $\beta$, increasing the concentration by a given amount $\Delta c_{\rm gNFW}$ leads to a larger increase in $\rho_0$ compared to the same concentration increase for a subhalo with a larger value of $\beta$. Therefore, for models fit with a lower $\beta$, stronger constraints on concentration can be obtained than for models with larger $\beta$. This can be seen from the blue slope-concentration contour, which is narrower for smaller $\beta$ and widens as $\beta$ increases. The gNFW profile is a (smoothly) broken power-law with a fixed outer slope of 3, so in the limit where $\beta \to 3$, the parameter $c_{\rm gNFW}$ becomes completely meaningless and therefore cannot be constrained at all regardless of the $\beta$ posterior. This is evident as the red slope-concentration contour is significantly flatter than the blue and extends over a wider concentration range. Interestingly, this contour does not extend across the full concentration prior range in the case of the $10^{10}$~M$_\odot$ subhalo. This suggests that something is breaking the degeneracy between inner slope and concentration and ruling out exceedingly high concentrations.

To demonstrate how the parameter inference scales with subhalo mass, the right panel of \figref{fig:HST_OnRing_cornerplots} shows the results for the Steep and NFW subhalos with a mass closest to the minimum detectable mass for each slope. This is determined by finding the subhalo with a mass closest to (but greater than) the point where each curve of \figref{fig:detectability} crosses the detectability threshold. In this case, the subhalo masses are inferred to be $\log_{10}(M_{\rm 200, Steep}/$M$_\odot)=8.9^{+0.2}_{-0.1}$ (truth: $8.8$) and $\log_{10}(M_{\rm 200, NFW}/$M$_\odot)=9.9^{+0.3}_{-0.3}$ (truth: $9.9$). The Steep subhalo's inner slope is inferred to be $\beta_{\rm Steep}=2.03^{+0.25}_{-0.51}$, which is not as stringent as for the $10^{10}$~M$_\odot$ case. Given this posterior, there is a 13\% (41\%) chance of inferring an inner slope $< 1.5\ (2)$ for this subhalo.  The inner slope of the NFW subhalo is inferred to be $\beta_{\rm NFW}=0.82^{+0.44}_{-0.52}$. The probability of mistaking this NFW subhalo for a Steep subhalo is still very unlikely with only a $3\%$ chance of inferring an inner slope $>2$. The slope-concentration correlation explained above can be seen here as well. The blue NFW contour is still banana-shaped as before, and now the Steep subhalo's slope-concentration contour also has this shape. This is because the true subhalo mass is lower and hence the strength of the subhalo's perturbation is weaker than for the $10^{10}$~M$_\odot$ subhalo. Therefore, constraints are weaker overall, even for the Steep subhalo. However, the overall trend of better concentration constraints for a lower-$\beta$ subhalo still holds with concentration constrained to be $c_{\scriptscriptstyle{\rm gNFW, NFW}}=15.6^{+11.6}_{-8.8}$ for the NFW subhalo (truth: 11.9) and $c_{\scriptscriptstyle{\rm gNFW, Steep}}=22.4^{+46.3}_{-18.7}$ for the Steep subhalo (truth: 13.8).

\begin{figure*}
    \centering
    \includegraphics[scale=0.6]{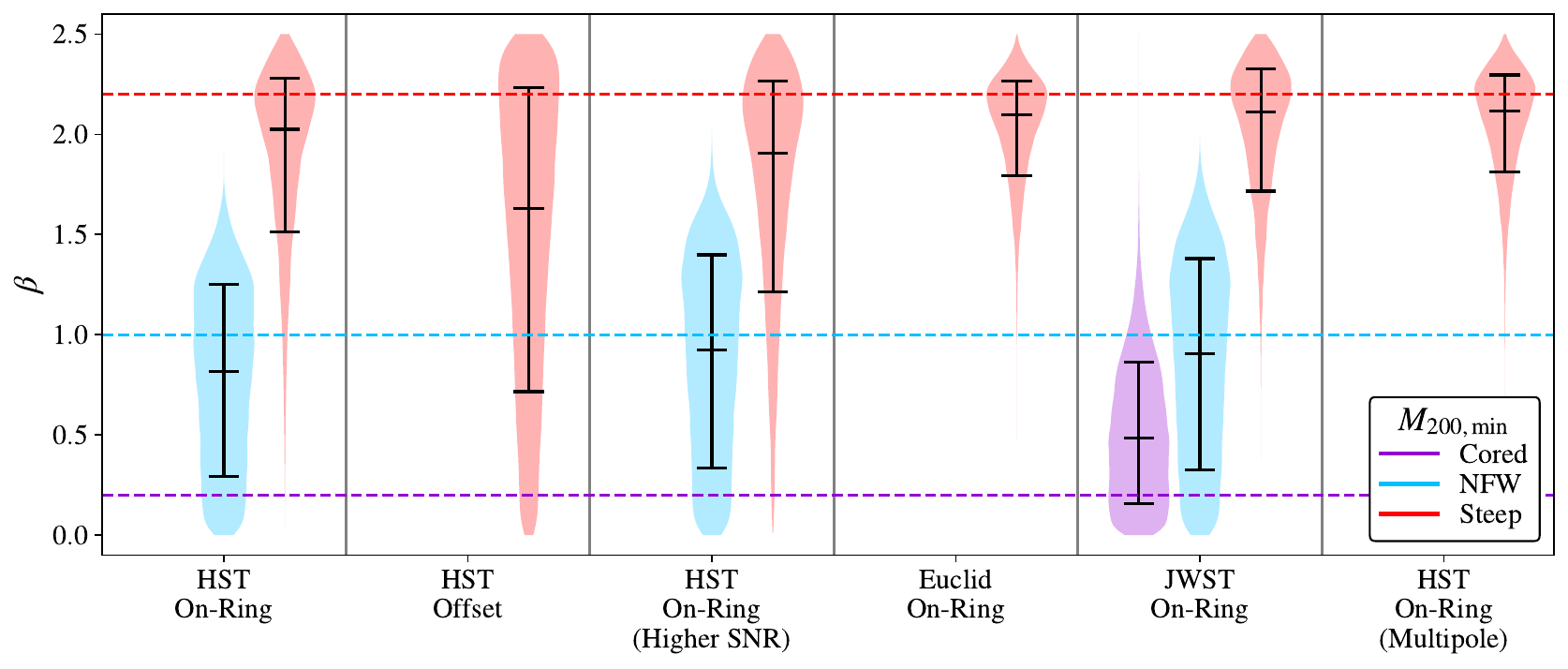}
    \includegraphics[scale=0.6]{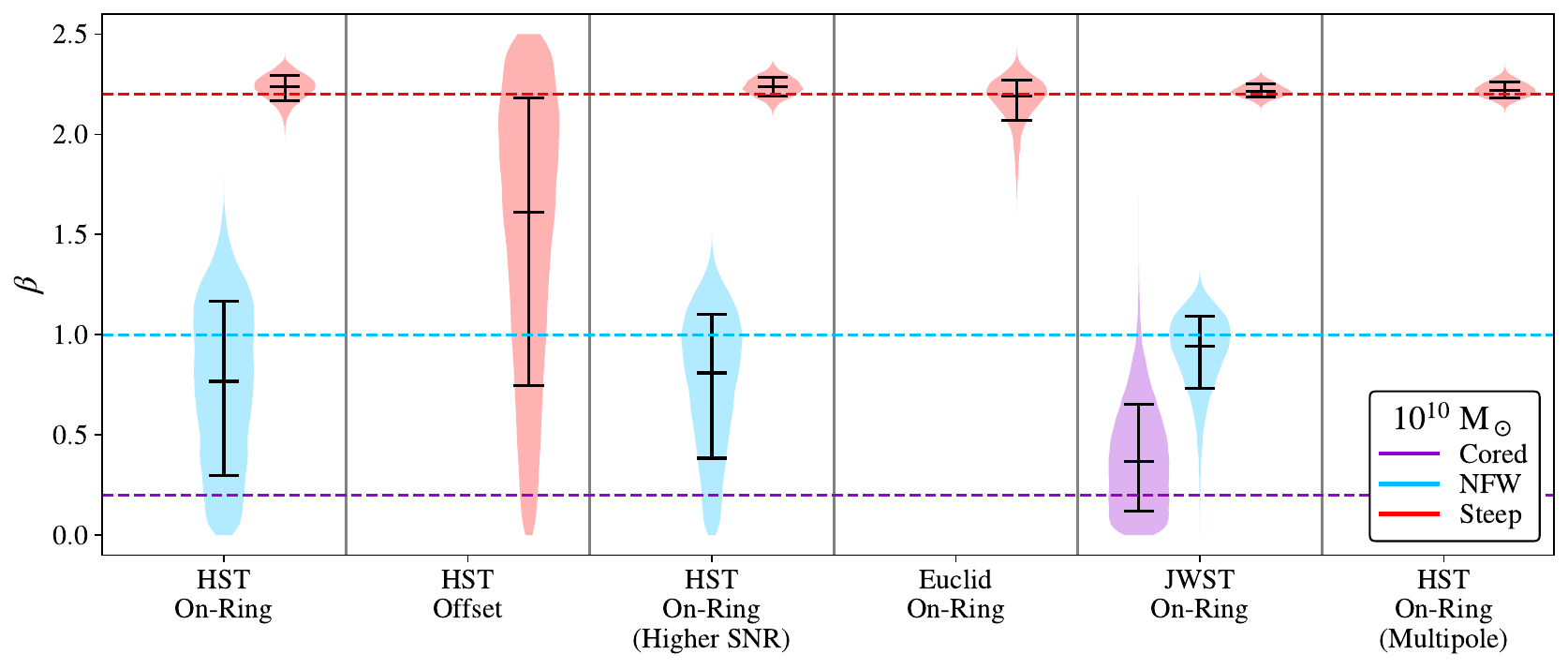}
    \caption{Constraints on the inferred value of $\beta$ for the subhalo with a mass closest to the minimum value detectable for each variation and each inner slope~(top) and for the $10^{10}$~M$_\odot$ subhalos~(bottom). The black lines bracket the 68\% confidence region with the middle line corresponding to the median. Constraints are not shown for any subhalos that do not meet the detectability threshold. $\beta$ is robustly inferred for the Steep subhalos across all variations except when the subhalo is offset from the Einstein ring. Most notably, $\beta$ is still strongly constrained even when multipoles are added to the lens model. Informative constraints on $\beta$ are obtained for the NFW subhalos in the HST On-Ring, HST On-Ring~(Higher SNR), and JWST On-Ring suites. Constraints on $\beta$ for the Cored subhalos are obtained only for the JWST On-Ring suite.}
    \label{fig:beta_constraints}
\end{figure*}

The results of all five suites are summarized in Figs.~\ref{fig:M_lowest} and~\ref{fig:beta_constraints}. For each variation, \figref{fig:M_lowest} shows the minimum-mass subhalo that is detectable (corresponding to $\Delta \ln \varepsilon=50$) for Cored~(purple), NFW~(blue), and Steep~(red) subhalos. This mass is interpolated from the $\Delta\ln\varepsilon$ versus true subhalo mass curves, like those shown in \figref{fig:detectability}. An upward arrow indicates when no subhalos are detectable for a specific variation up to the maximum mass of $10^{10}$~M$_\odot$. The violin plots of \figref{fig:beta_constraints} illustrate how well $\beta$ is constrained for the minimum-mass detectable subhalo (top) and the $10^{10}$~M$_\odot$ subhalo (bottom). The horizontal dashed lines mark the true $\beta$ value for the corresponding colored data points. The black lines on each contour mark the 68\% confidence region on the inferred value of $\beta$ for that subhalo. Note that  \figref{fig:beta_constraints} does not show any $\beta$ constraints for subhalos that do not meet the detectability threshold. Supplementary figures in the Appendix complement the results presented in the following subsections. \figref{fig:detectability_plots} provides plots of $\Delta \ln \varepsilon$ versus true subhalo mass for the suites discussed in these subsections. Figs.~\ref{fig:HST_Offset_cornerplots}--\ref{fig:HST_OnRing_m134_cornerplots} provide corner plots of the posterior probability distributions for the $10^{10}$~M$_\odot$ and minimum-mass-detectable subhalos of these suites.

\subsection{Impact of Subhalo Position} \label{sec:offset}

When located on the Einstein ring, Steep subhalos are significantly more detectable than Cored and NFW subhalos. This is consistent with expectations, as the inner slope of a subhalo's density distribution impacts how it perturbs nearby source emission---see the right panel of \figref{fig:def_curve_dens}, which shows that for small $R$, the subhalo's deflection strength grows with the steepness of its inner slope. However, as $R$ increases, the difference in deflection strength between subhalos of various inner slopes decreases and eventually becomes roughly independent of $\beta$. This subsection considers the HST Offset suite, where the subhalo is positioned off the Einstein ring with a projected position in the lens plane of (0.3\arcsec, 2.3\arcsec). In this case, the nearest source emission that the subhalo perturbs is $\sim1$\arcsec~away. With no source emission in the immediate vicinity of the subhalo, the maximal amount that a Steep subhalo can perturb the source emission decreases and is closer in value to that of the Cored and NFW subhalos. Therefore, the detectability of Offset subhalos should depend less on the  inner slope. 

\figref{fig:M_lowest} shows that the lowest mass detectable for the Steep subhalos increases from $5.4\times10^8$~M$_\odot$ to $3.2\times10^9$~M$_\odot$ when moved off the Einstein ring. Additionally, all NFW and Cored subhalos fall below the detectability threshold. The top left panel of \figref{fig:detectability_plots} shows that the Steep subhalo curve shifts closer to the NFW one, indicating that Steep subhalos are no longer significantly more detectable than their NFW counterparts. Interestingly, as seen from \figref{fig:detectability_plots}, two Cored subhalos now have a $\Delta \ln \varepsilon >5$ when moved off the Einstein ring. This is likely a result of how the Cored subhalo deflection strength, $\alpha(R)$, peaks at a distance away from the subhalo's location, as can be seen in \figref{fig:def_curve_dens}. The second column of each \figref{fig:beta_constraints} panel shows that there is a clear and significant loss in constraining power on $\beta$ for detectable Steep subhalos in this suite relative to the HST On-Ring results. The $\beta$ constraints for the $10^{10}$~M$_\odot$ Steep subhalo rule out inner slopes $< 2$ for the HST On-Ring suite. For the HST Offset suite, there is a $68$\% probability for this subhalo to have an inner slope $< 2$. These results highlight the importance of subhalo position for detectability and parameter constraints.

\subsection{Impact of Data Quality} \label{sec:quality_results}

This subsection presents the results for  the HST~(Higher-SNR), Euclid, and JWST On-Ring suites, which differ from the HST On-Ring suite in regards to data quality. The true subhalo position in these suites corresponds to the On-Ring position, with a projected location of $(0.3'',~1.3'')$ on the Einstein ring. The HST On-Ring~(Higher-SNR) analysis investigates how doubling SNR$_{\rm max}$ from $\sim30$ to $\sim60$ impacts the minimum-mass subhalo detectable in the data and the inference of subhalo properties. For this suite, the minimum detectable mass decreases to $2.3\times10^8$~M$_\odot$ and $1.2\times10^9$~M$_\odot$ for Steep and NFW subhalos, respectively. While the $\Delta \ln \varepsilon$ of the Cored subhalos also increases in this case, they still fall below the detectability threshold. Overall, there are slight improvements in constraining power for this case. For the $10^{10}$~M$_\odot$ subhalos, the inner slope is inferred to be $\beta_{\rm Steep}=2.24^{+0.05}_{-0.05}$ for the Steep subhalo and $\beta_{\rm NFW}=0.81^{+0.29}_{-0.42}$ for the NFW subhalo.

With the massive number of strong lensing observations available from Euclid in the near future, it is important to investigate how well subhalos can be detected and their properties inferred from these observations. For the Euclid On-Ring suite, Cored and NFW subhalos with masses $\leq 10^{10}$~M$_\odot$ are not detectable, but Steep subhalos are detectable down to $2.2\times10^9$~M$_\odot$, as summarized in \figref{fig:M_lowest}. The inner slope of these detectable subhalos is well-constrained, with $\beta_{\rm Steep}=2.19^{+0.08}_{-0.12}$ for the $10^{10}$~M$_\odot$ subhalo and $\beta_{\rm Steep}=2.10^{+0.17}_{-0.30}$ for the minimum-mass detectable subhalo. 

For the JWST On-Ring suite, the minimum detectable mass for the Steep and NFW subhalos decreases down to $8.3\times10^7$~M$_\odot$ and $4.5\times10^8$~M$_\odot$, respectively. Additionally, Cored subhalos, which are undetectable in HST- and Euclid-like data, are now detectable down to masses of $2.7\times10^9$~M$_\odot$. The strongest constraints on all three subhalo parameters are obtained with JWST-like data. For example, the inner slopes of the $10^{10}$~M$_\odot$ subhalos are inferred to be $\beta_{\rm Steep}=2.22^{+0.04}_{-0.03}$, $\beta_{\rm NFW}=0.94^{+0.15}_{-0.21}$, and $\beta_{\rm Cored}=0.37^{+0.29}_{-0.25}$ for the Steep, NFW, and Cored subhalos, respectively.

\subsection{Impact of Multipoles in Lens Model} \label{sec:mult_results}
When analyzing a real strong lens system, it is often necessary to include multipole perturbations in the lens model to account for the complexity of the main lens galaxy's mass distribution. We re-model all HST On-Ring and Offset observations (with SNR$_{\rm max}\sim30$) to investigate how adding 1st-, 3rd-, and 4th-order multipoles to the model impacts subhalo detectability. When multipoles are added to the model, all previously undetectable subhalos in the HST On-Ring and Offset data remain undetectable. Additionally, all On-Ring NFW and Offset Steep subhalos---which were previously detectable---now fall below the threshold. Remarkably, however, Steep subhalos along the Einstein ring remain highly detectable in the data: the minimum-mass Steep subhalo detectable in the HST On-Ring suite increases only slightly to $8.3\times10^8$~M$_\odot$ when multipoles are added. This highlights the uniqueness of the Steep subhalo's lensing signal compared to that of Cored subhalos, NFW subhalos, and multipole perturbations.

The $\beta$ posteriors for the Steep On-Ring minimum-mass detectable subhalo and the $10^{10}$~M$_\odot$ subhalo are shown in the top and bottom panels of \figref{fig:beta_constraints}, respectively. The bottom panel of \figref{fig:beta_constraints} suggests that $\beta$ is better constrained for the $10^{10}$~M$_\odot$ Steep subhalo when multipoles are included in the model~(last column) as opposed to when they are not~(first column). Specifically, the inner slope is inferred to be $\beta_{\rm Steep}=2.22^{+0.04}_{-0.04}$ when multipoles are included, compared to $\beta_{\rm steep}=2.24^{+0.06}_{-0.07}$ when they are not. The narrowing of the posteriors is surprising, given the additional free parameters in the multipole model.  By comparing the left panel of \figref{fig:HST_OnRing_cornerplots} with that of \figref{fig:HST_OnRing_m134_cornerplots} it can be seen that the joint posterior distributions for the $10^{10}$~M$_\odot$ Steep subhalo shrink when multipoles are added. However, when multipoles are included, the joint posterior distributions are not as well centered around the true parameter values compared to the case with no multipoles.

Adding multipoles to the lens model adds six additional free parameters, which increases the dimensionality of the parameter space that \texttt{Nautilus} searches. The difference in parameter constraints when multipoles are added may suggest that \texttt{Nautilus} has converged on a local, as opposed to global, maximum. As a quick check of this hypothesis, we doubled the number of live points \texttt{Nautilus} uses for the Lens Mass Pipeline and Subhalo Pipeline runs for the $10^{10}$~M$_\odot$ Steep subhalo in the HST On-Ring~(Multipole) suite, but the results remained nearly unchanged. We save a more detailed study of these biases in the parameter recovery for future work.

\section{Results with a Pixelized Source Reconstruction}
\label{sec:pixelized}

The results presented so far provide intuition for how a subhalo's detectability depends on its inner density slope and position, as well as the data quality and complexity of the main lens mass model. The fitting procedure for this analysis treated the source galaxy’s light using a cored-Sérsic profile, as was originally used to  simulate the data. Using this parametric form reduced the complexity of the modeling procedure and the computational time. However, when analyzing real strong lensing observations (which can have complex source galaxy morphologies), a pixelized source reconstruction is often used \citep{2014MNRAS.442.2017V, 2024MNRAS.52710480N}. This procedure divides the  source plane into an irregular grid of pixels and the free parameters describing the source galaxy's light distribution are the pixel fluxes. While such a model is important for capturing any irregular aspects of the source galaxy's morphology, its added flexibility can make detecting a subhalo in the lens system more difficult. This is because the subhalo's lensing effect can be degenerate with reasonable changes to the source galaxy's morphology. 

A natural next step is to repeat this study using a pixelized source reconstruction to investigate how the additional model flexibility impacts the results found in \secref{sec:results}. A full analysis that looks at all variations considered here (i.e., subhalo properties, data quality, and the complexity of the main lens mass model) but with a pixelized source reconstruction is beyond the scope of this work. However, as an initial test, we re-model four HST On-Ring mock observations (those with an NFW or Steep subhalo with $\log_{10}(M_{200}/$M$_\odot) = 9.1$ and 9.5) using a pixelized source reconstruction. We do not re-model the observations with a Cored subhalo, as those correspond to $\Delta\ln\varepsilon<5$ in \secref{sec:results}. We additionally fit the lens galaxy's light instead of subtracting it from the data. Below, we briefly describe how the new modeling procedure differs from that described in \secref{sec:fitting}. The complete formalism for pixelized source reconstruction and lens-light fitting is described in \cite{2024MNRAS.532.2441H} and \cite{2024MNRAS.52710480N}, and is used by \cite{2025arXiv250116139W} for substructure analysis of mock Euclid strong lenses. 

The modeling procedure performed here follows the non-linear search chaining approach used previously, with the addition of the Pixelized Source Pipeline and the Lens Light Pipeline from the SLaM series discussed in \secref{sec:fitting}. The modeling procedure begins with the Parametric Source Pipeline run as before but with two additional cored-Sérsic profiles to model the lens galaxy's light. The Parametric Source Pipeline initializes the main lens galaxy's mass model before running the subsequent Pixelized Source Pipeline, which is important to avoid inferring a demagnified solution~\citep{2021MNRAS.503.2229M}. These are solutions in which the lens galaxy has a very low mass and essentially does not lens the source light. By initializing priors for the lens galaxy's mass model based on the Parametric Source Pipeline results, \texttt{Nautilus} can be guided to the correct region of parameter space for the mass model.

For the Pixelized Source Pipeline, we use \texttt{PyAutoLens}’s flexible adaptive Voronoi-mesh source reconstruction. The mass parameters of the macro model are used to ray trace image-plane pixel coordinates to the source plane, where the Voronoi mesh is constructed. The source pixel fluxes are solved for via a linear inversion~\citep{Bro1997, 2003ApJ...590..673W, 2024MNRAS.532.2441H}. We adopt a Voronoi-mesh grid with Natural Neighbour interpolation~\citep{Sibson1981}. As the linear system is ill-posed, the source reconstruction is regularized using the bespoke adaptive regularization scheme described in Appendix~A of \cite{2024MNRAS.532.2441H}, which penalizes models with large source luminosity gradients, thereby favoring smoother source light distributions.  From this point on, the source galaxy is modeled using the pixelized source reconstruction as opposed to a parametric profile. Next, the Lens Light Pipeline is performed, which fixes the lens' mass based on the previous pipeline and re-models the lens galaxy's light. From this point on, the lens light model is held fixed based on the results of this pipeline. Then the Lens Mass Pipeline is run to fit the Fiducial mass model to the data, which uses an EPL profile to model the main lens galaxy's mass distribution. Finally, the Subhalo Pipeline is run as before, but with the source now modeled using the pixelized method. 

\tabref{tab:pix_results} lists the $\Delta\ln\varepsilon$ values obtained from this new modeling procedure and the original values obtained for these four HST On-Ring data sets. By comparing these values, one can assess how using a Voronoi source reconstruction and including lens light in the modeling impacts subhalo detectability. As before, we find that Steep subhalos are significantly more detectable than their NFW counterparts of the same mass. However, there is a clear decrease in $\Delta\ln\varepsilon$ for all four data sets when using the new modeling procedure. The $10^{9.1}$ and $10^{9.5}$~M$_\odot$ NFW subhalos both fall below the detectability threshold when using either modeling procedure. Previously, the $10^{9.1}$ and $10^{9.5}$~M$_\odot$ Steep subhalos were both detectable with $\Delta\ln\varepsilon$ values of 199 and 832, respectively. Using the more sophisticated modeling approach on these same data sets, both subhalos remain detectable, however, their $\Delta\ln\varepsilon$ values decrease to 88 and 251. 
We find little change to the inferred subhalo properties for these two detectable subhalos. For instance, with the original parametric source modeling procedure, the inner slope was inferred to be $\beta_{\rm Steep}=2.17^{+0.14}_{-0.26}$ and $\beta_{\rm Steep}=2.23^{+0.08}_{-0.11}$ for the $10^{9.1}$ and $10^{9.5}$~M$_\odot$ Steep subhalos, respectively. With the more sophisticated modeling procedure, the inner slope constraints for these two subhalos are $\beta_{\rm Steep}=2.11^{+0.22}_{-0.35}$ and $\beta_{\rm Steep}=2.16^{+0.15}_{-0.23}$. This preliminary test suggests that even with a pixelized source reconstruction and lens light fitting, Steep subhalos remain detectable and their true properties can be robustly recovered.

\section{Discussion and Conclusions} \label{sec:discussion_conclusions}
This paper investigated how the detectability of a dark matter subhalo in a strong lens system depends on its inner density slope, $\beta$. To do so, we simulated mock strong gravitational lensing observations consisting of a Cored $(\beta=0.2)$, NFW $(\beta=1)$, or Steep $(\beta=2.2)$ subhalo perturber. The Cored and Steep profiles considered in this work arise in SIDM, a particularly well-motivated case study, as well as in other non-CDM variants. For detectable subhalos, we investigated how robustly their properties were inferred from the data and the accuracy to which the various subhalo profiles could be distinguished from one another. 

\begin{table}
 \renewcommand{\arraystretch}{1.2} 
 \begin{tabular*}{\columnwidth}{@{}l@{\hspace*{20pt}}l@{\hspace*{25pt}}l@{\hspace*{15pt}}l@{}}
  \toprule
  \multicolumn{2}{l}{\textbf{HST On-Ring Obs.}} & \multicolumn{2}{c}{$\bm{\Delta\ln\varepsilon}$} \\
  \textbf{Slope} & $\bm{\log_{10}(m_{200})}$ & \textbf{Parametric Source} & \textbf{Pixelized Source} \\
  \midrule
  NFW & 9.1 & 6 & 2 \\
  NFW & 9.5 & 24 & 6 \\
  Steep & 9.1 & 199 & 88 \\
  Steep & 9.5 & 832 & 251 \\
  \bottomrule
   \end{tabular*}
   \caption{$\Delta\ln\varepsilon$ values obtained for four HST On-Ring data sets from different modeling procedures. The third column lists the values presented in \secref{sec:results} which were obtained using the modeling procedure described in \secref{sec:fitting} with a parametric source model. The fourth column lists the values obtained when using a pixelized source reconstruction and fitting the light of the main lens galaxy, as described in \secref{sec:pixelized}.}
 \label{tab:pix_results}
\end{table}

The key result of this paper is that subhalo detectability in a strong lens system depends critically on the subhalo's inner slope, with Steep subhalos being significantly more detectable than their NFW or Cored counterparts. The benchmark suite consisted of HST-quality data representative of existing surveys \citep[e.g., SLACS,][]{2005ApJ...624L..21B}, with a subhalo perturber positioned along the Einstein ring of the system. With the analysis procedure described in \secref{sec:fitting}, Steep subhalos are detectable ($\Delta \ln \varepsilon\geq50$) down to $M_{200} \sim5.4\times10^{8}$~M$_\odot$. In contrast, the mass threshold for detecting NFW subhalos is over an order of magnitude higher, while Cored subhalos are undetectable up to at least $10^{10}$~M$_\odot$. For detectable subhalos in this suite, $\beta$ was robustly inferred, with constraints improving for more massive systems or steeper profiles. When the subhalo was moved off the Einstein ring in this HST-quality data, the sensitivity to the inner slope decreased significantly, as expected.

Previous works \citep[e.g.,][]{2021MNRAS.507.1202M, Amorisco2022:2022MNRAS.510.2464A} showed that subhalo detectability scales with the concentration of the subhalo. While this result is similar to what is found here, we emphasize that such studies did not consider profiles as diverse as those considered in this work. The Steep profile used here to represent core-collapsed subhalos results in lensing effects significantly stronger than what can be achieved with a standard NFW profile. For an NFW subhalo to have a lensing effect even remotely similar to that of a Steep subhalo, an absurdly high concentration, significantly greater than what is predicted by CDM, would be necessary. Furthermore, our results indicate that as expected, a strong degeneracy exists between concentration and $\beta$. However, between these two parameters, $\beta$ is more robustly inferred.

When analyzing real data, a pixelized source reconstruction is often used, as source galaxies can have complex morphologies that cannot be properly modeled by a parametric profile. However, using this more flexible source galaxy model can significantly reduce subhalo detectability, as the subhalo's lensing effect can be degenerate with changes to the source galaxy's morphology. In \secref{sec:pixelized}, we re-analyzed four observations from the HST On-Ring suite (those with an NFW or Steep subhalo with $\log_{10}(M_{200}/$M$_\odot) = 9.1$ and 9.5) using a pixelized source reconstruction to verify that the main results of this work continue to hold. Importantly, the Steep subhalos remained significantly more detectable than their NFW counterparts, despite the level of detectability decreasing in all cases. Additionally, the properties of the Steep subhalos were robustly inferred in both cases. 

One of the most challenging aspects of strong lensing subhalo studies is the degeneracy between macro-model complexity and subhalo perturbations. Previous studies have shown that using an overly simplistic mass model for the main lens galaxy can result in false-positive subhalo detections~\citep{He:2023MNRAS.518..220H, ORiordan:2024MNRAS.528.1757O} that attempt to account for the unmodeled lens complexity. On the other hand, including additional complexity in the macro model (e.g., via multipoles) can absorb subhalo lensing effects and substantially reduce  detectability~\citep{ORiordan:2024MNRAS.528.1757O, 2025arXiv250902660O, 2025MNRAS.539..704L}. These previous investigations focused on subhalos with NFW density profiles. Importantly, our main results (using the modeling procedure described in \secref{sec:fitting}) demonstrate that subhalos with sufficiently steep inner slopes that are located on the Einstein ring break this degeneracy with the macro model. Including 1st-, 3rd-, and 4th-order multipole perturbations in the lens-mass model strongly reduces the detectability of Cored and NFW subhalos in the HST On-Ring suite to $\Delta\ln\varepsilon<5$. By contrast, the detectability of Steep subhalos in this suite decreases only slightly when multipoles are added to the model, with masses down to $\sim8.3\times10^{8}$~M$_\odot$ remaining detectable. Note that this protection from degeneracy does not hold when the subhalo is sufficiently offset from the Einstein ring, as the few Steep subhalos detectable in the HST Offset suite drop below the detectability threshold when multipoles are included in the model.

These results are intuitive when considered in terms of deflection angles. A Steep subhalo has deflection angles that vary rapidly on $\sim$1~kpc scales, whereas standard prescriptions for increasing macro-model complexity---whether multipoles or models that separately fit stars and dark matter \citep[e.g.,][]{2019MNRAS.489.2049N}---do not generate such rapidly varying deflections. While such macro-model complexity is strongly degenerate with shallower subhalo perturbations, we find that even at extreme parameter values, multipole perturbations struggle to mimic the lensing signature of Steep subhalos positioned near the Einstein ring. Therefore, our results indicate that subhalos with sufficiently steep inner slopes are not only more detectable, \textit{but may also be easier to distinguish from the most problematic macro-model systematics in strong lens analyses}. In this work, we have separately demonstrated that Steep subhalos along the Einstein ring remain very detectable when multipoles are included in the lens model and when a pixelized source reconstruction is used. The natural next step is to bring these two modeling approaches together to investigate subhalo detectability when simultaneously applying complex lens and source modeling techniques.

This work highlights how subhalos with steep inner density profiles are significantly more detectable than those with shallower profiles. The number of subhalo detections expected from analyzing a set number of strong lenses is highly dependent on the dark matter model, with more detections expected from models that produce an abundance of very concentrated substructures. Current subhalo detections from strong lensing observations may already be evidence of this selection effect, as three of the detected subhalos have been shown to be well fit by steep density profiles with inner slopes that are close to or steeper than isothermal \citep{2021MNRAS.507.1662M, 2025A&A...699A.222D, Minor2025, 2025MNRAS.540..247E, 2025MNRAS.543..540T, 2025arXiv251011006Y, 2025NatAs...9.1714P}. In the near future, the quantity and quality of strong lensing observations is expected to improve significantly as Euclid is on track to discover $\sim170{,}000$ strong lenses~\citep{2025arXiv250315325E, ORiordan2023}, and JWST enables imaging with unprecedented signal-to-noise~\citep{2025arXiv250308777N}. We additionally investigated the potential of both instruments for strong lensing subhalo detections and constraints. With mock Euclid-like data (lower resolution than HST), Steep subhalos down to $\sim2.2\times10^{9}$~M$_\odot$ remain detectable, while JWST-like data pushes detectability down to $\sim8.3\times10^{7}$~M$_\odot$ (when using a parametric source model). These results indicate that Euclid, JWST, and other upcoming facilities could provide even more powerful tests for dark matter constraints. This study does not address the expected abundance of detectable core-collapsed SIDM subhalos, as that is beyond the scope of this work. Performing full, quantitative forecasting is the natural next step, but the results here already highlight that strong lensing offers multiple, complementary avenues to constrain SIDM and other dark matter models that would lead to similarly concentrated substructures.

\section*{Acknowledgements}
We acknowledge Manoj Kaplinghat, Anna Nierenberg, Thomas Collett, Qiuhan He, and Aristeidis Amvrosiadis for helpful discussions. KEK is supported by the National Science Foundation Graduate Research Fellowship Program under Grant No. DGE-2444107. ML and KEK are supported by the Department of Energy~(DOE) under Award Number DE-SC0007968, as well as the Binational Science Foundation~(Grant Number~2022287).  ML is also supported by the Simons Investigator in Physics Award. This work was performed in part at Aspen Center for Physics, which is supported by National Science Foundation grant PHY-2210452. JWN is supported by an STFC/UKRI Ernest Rutherford Fellowship, Project Reference: ST/X003086/1. OS is supported by the Israel Science Foundation (grant No. 3104/25), and by the NSF-BSF (grant No. 2024829). The work presented in this article was performed on computational resources managed and supported by Princeton Research Computing, a consortium of groups including the Princeton Institute for Computational Science and Engineering~(PICSciE) and the Office of Information Technology's High Performance Computing Center and Visualization Laboratory at Princeton University.

\section*{Data Availability}

The data underlying this article will be shared on reasonable request to the corresponding author.



\input{main.bbl}

\appendix

\section{Supplementary Figures}
\label{sec:appendix}

\setcounter{equation}{0}
\setcounter{figure}{0} 
\setcounter{table}{0}
\renewcommand{\theequation}{A\arabic{equation}}
\renewcommand{\thefigure}{A\arabic{figure}}
\renewcommand{\thetable}{A\arabic{table}}

\begin{figure*}
    \centering
\includegraphics[width=0.9\columnwidth]{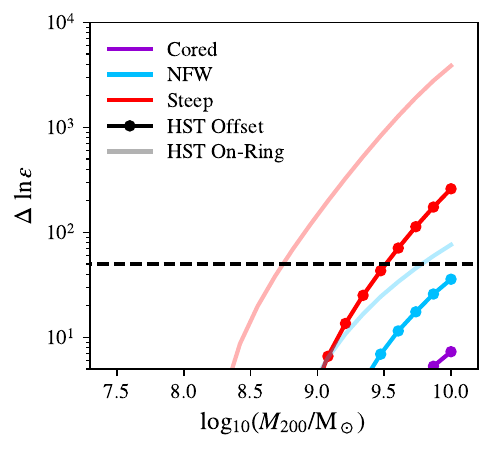}
    \includegraphics[width=0.9\columnwidth]{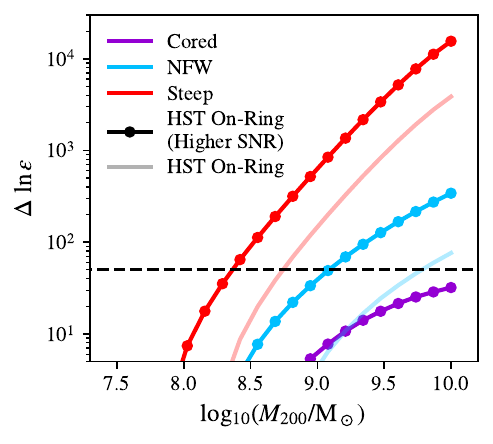}
    \includegraphics[width=0.9\columnwidth]{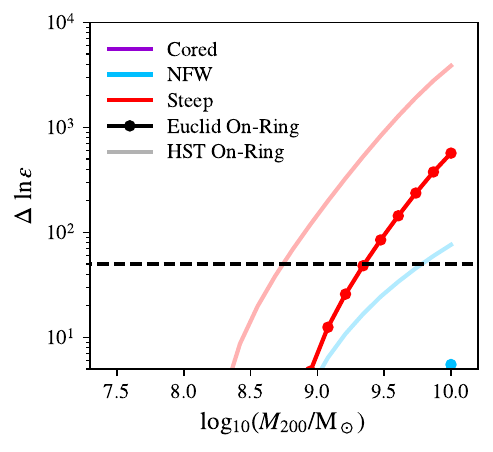}
    \includegraphics[width=0.9\columnwidth]{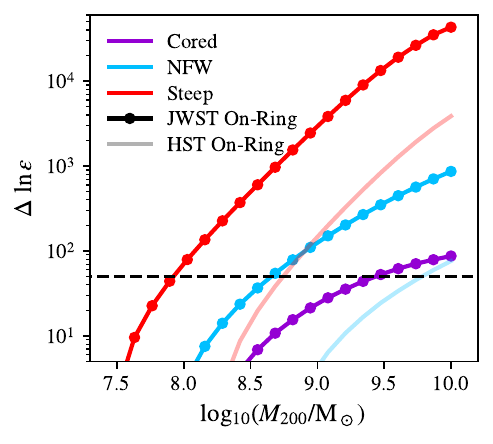}
    \includegraphics[width=0.9\columnwidth]{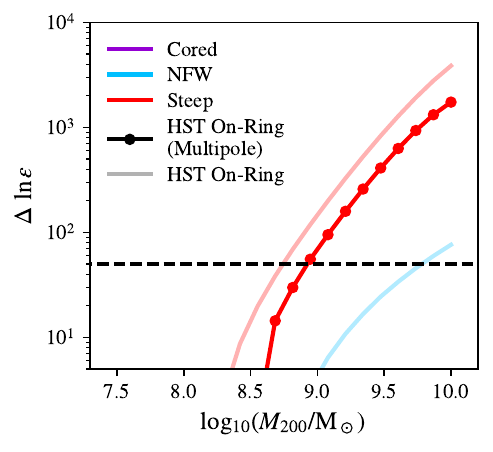}
    \includegraphics[width=0.9\columnwidth]{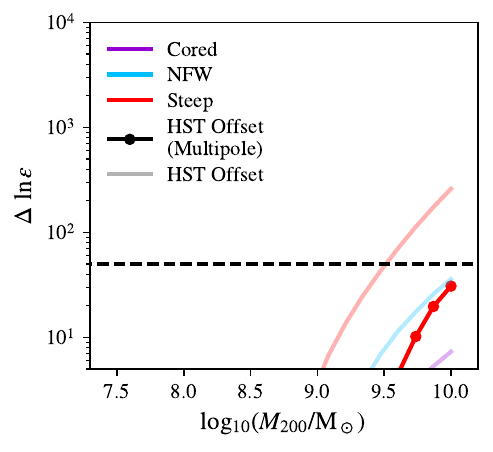}
\caption{Same as \figref{fig:detectability},
except for the following suites: HST Offset~(Row 1, left), HST On-Ring~(Higher SNR)~(Row 1, right), Euclid On-Ring~(Row 2, left), JWST On-Ring~(Row 2, right), HST On-Ring~(Multipole)~(Row 3, left), and HST Offset~(Multipole)~(Row 3, right). The HST On-Ring results from \figref{fig:detectability} are faintly plotted on each panel for comparison,  except for the HST Offset~(Multipole) plot, where the HST Offset results are plotted in the background instead.}    \label{fig:detectability_plots}
\end{figure*}

\begin{figure*}
    \centering
    \includegraphics[width=0.49\textwidth]{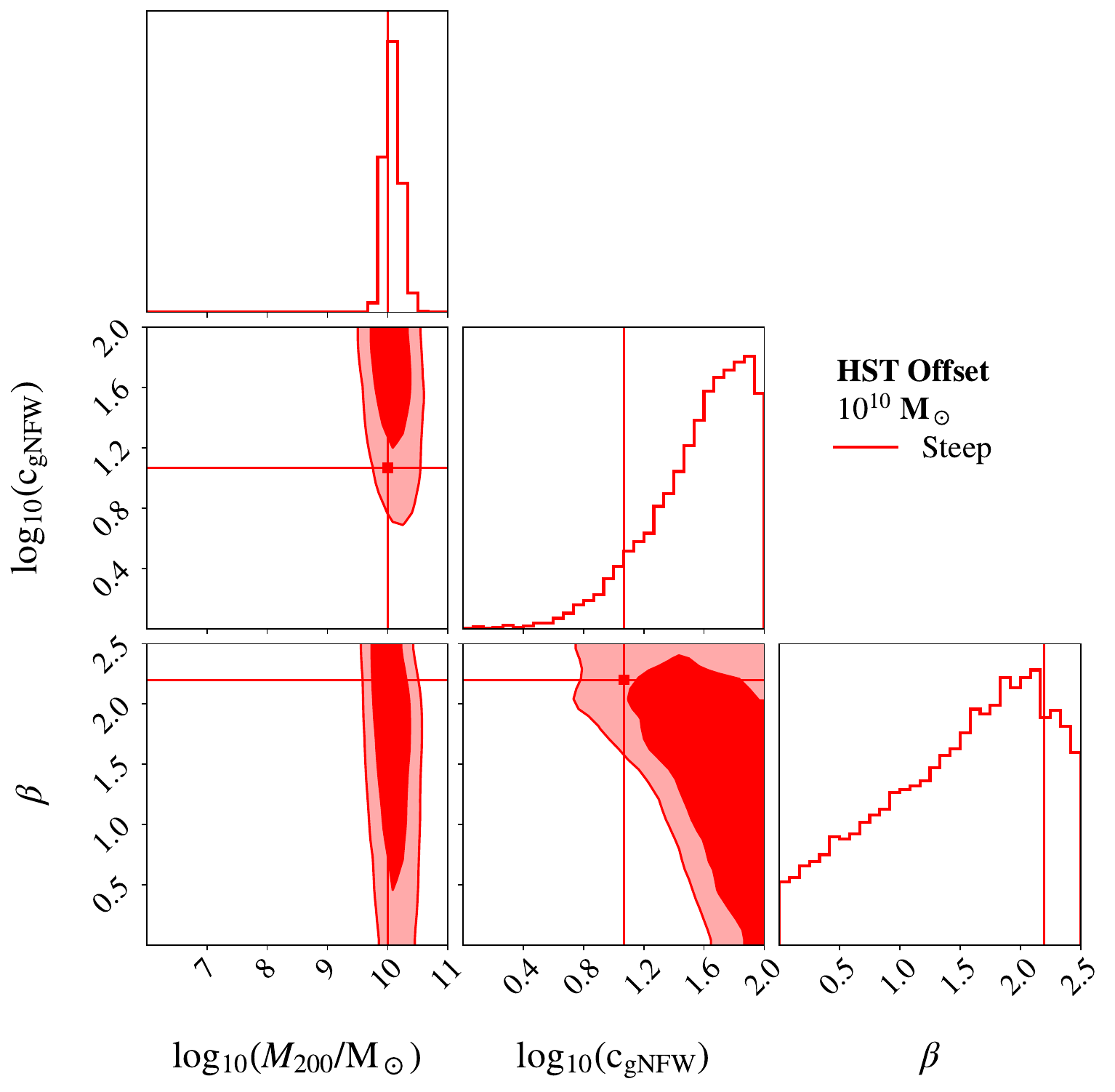}
    \includegraphics[width=0.49\textwidth]{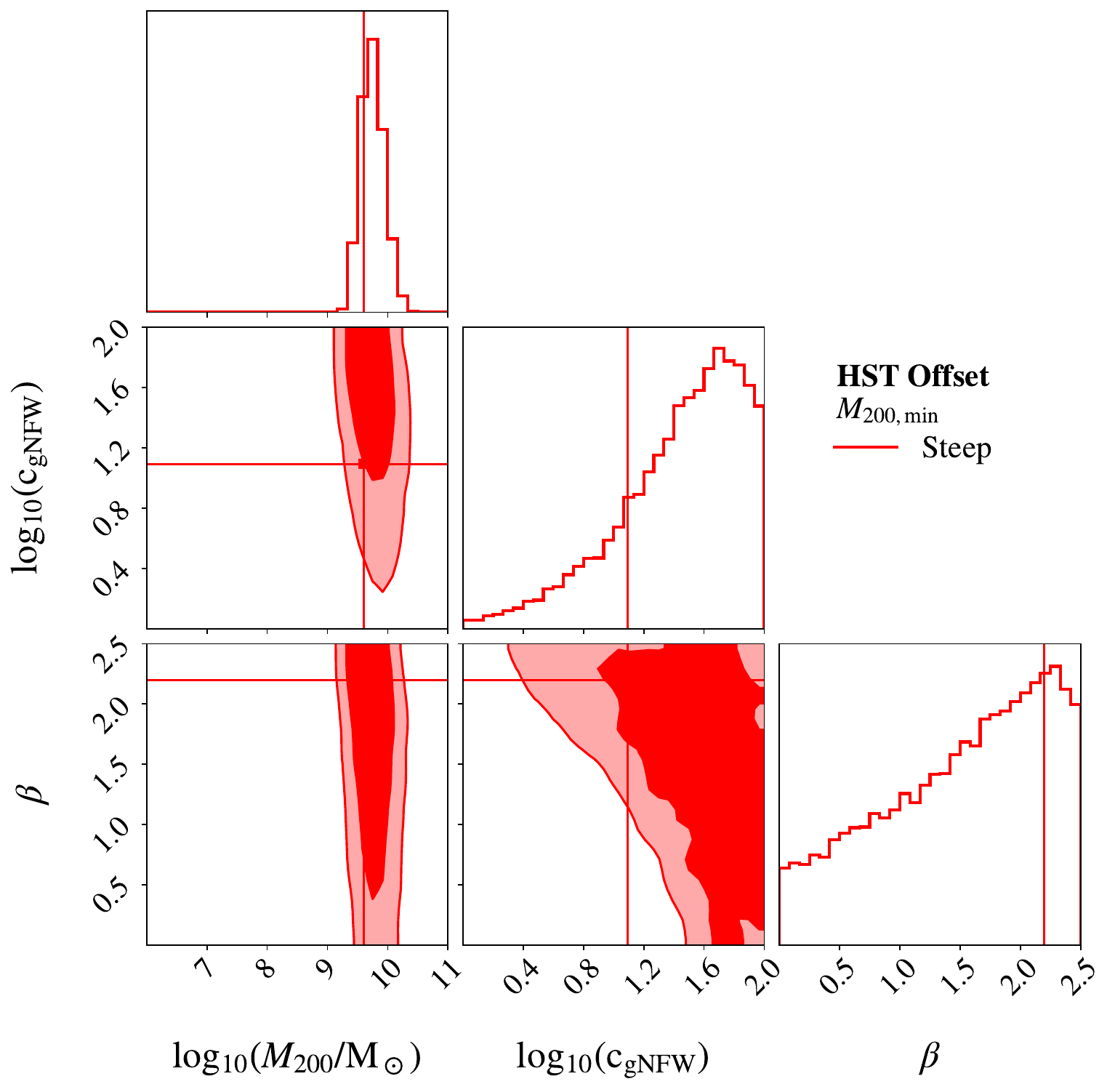}
    \caption{Same as \figref{fig:HST_OnRing_cornerplots}, except for the HST Offset suite.}
    \label{fig:HST_Offset_cornerplots}
\end{figure*}

\begin{figure*}
    \centering
    \includegraphics[width=0.49\textwidth]{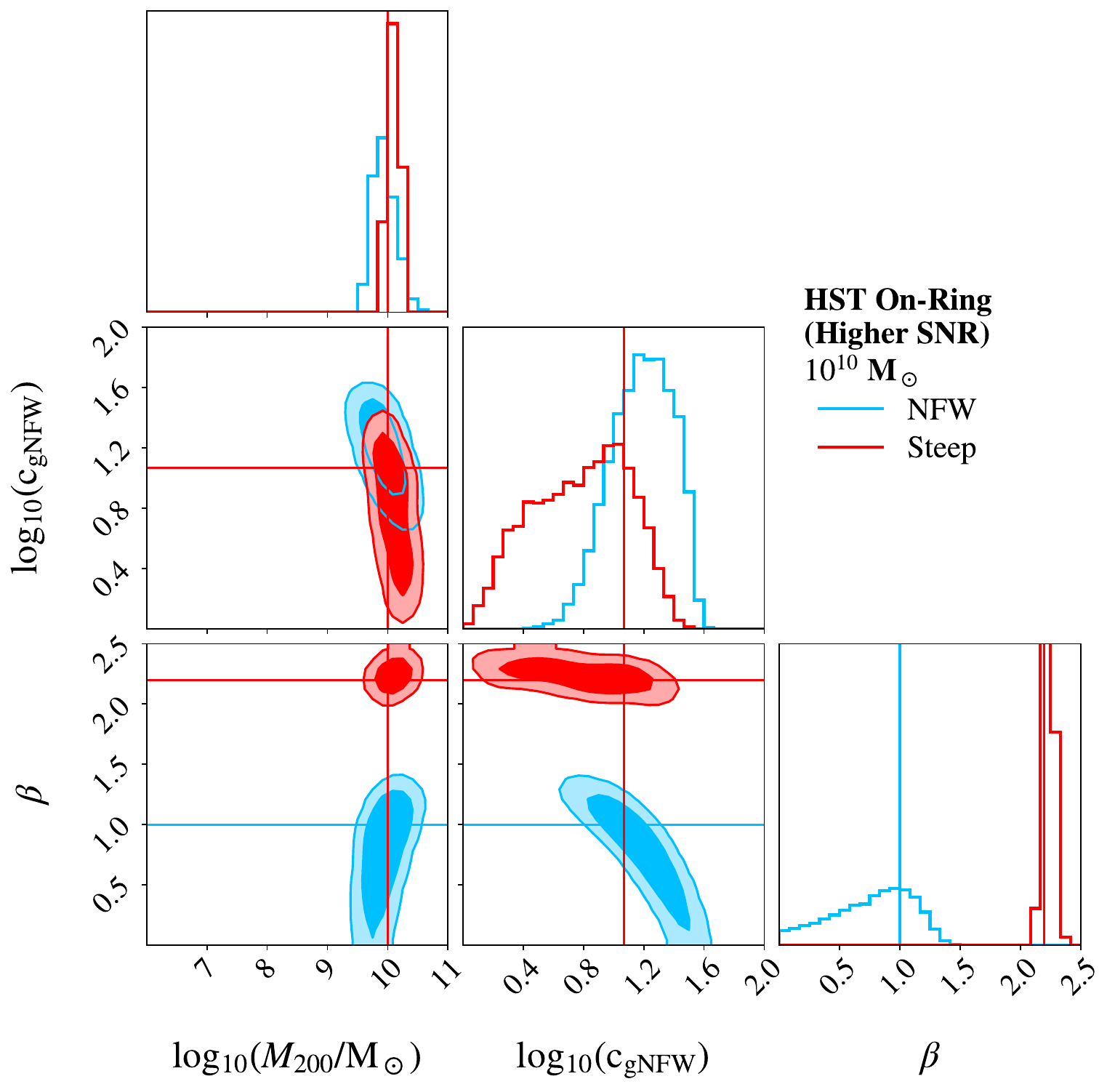}
    \includegraphics[width=0.49\textwidth]{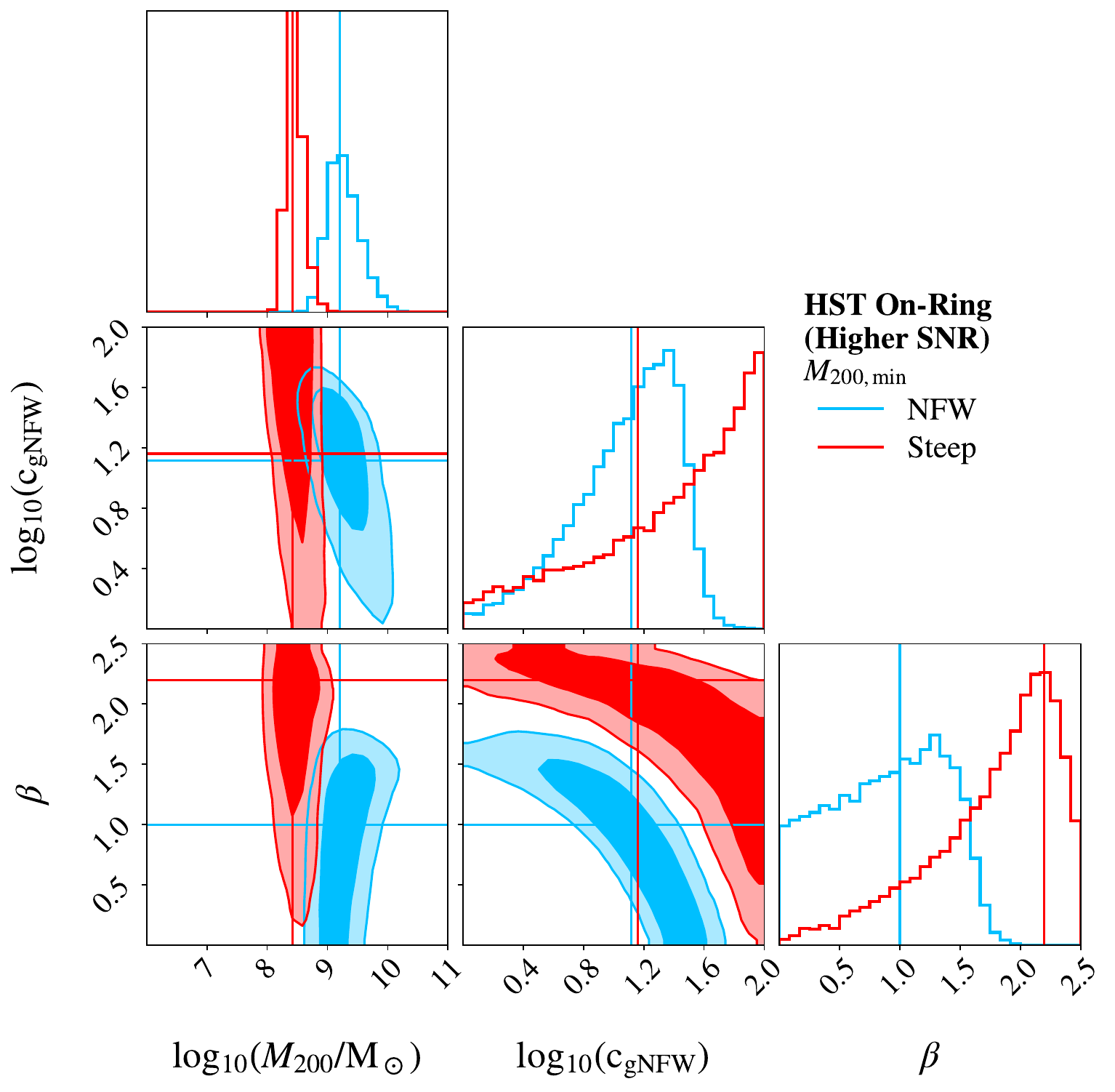}
    \caption{Same as \figref{fig:HST_OnRing_cornerplots}, except for the HST On-Ring (Higher SNR) suite.}
    \label{fig:HST_OnRing_HigherSNR_cornerplots}
\end{figure*}

\begin{figure*}
    \centering
    \includegraphics[width=0.49\textwidth]{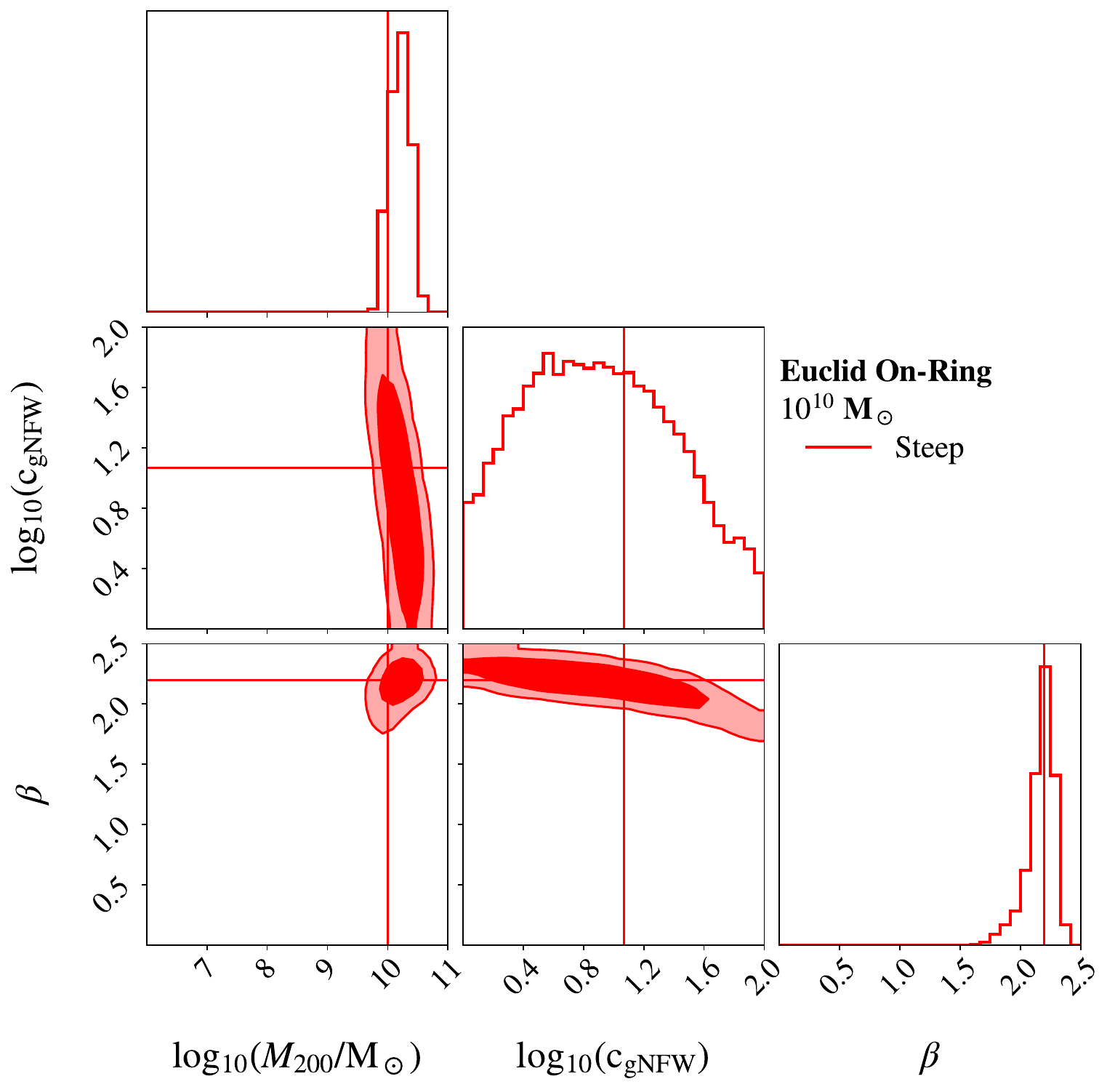}
    \includegraphics[width=0.49\textwidth]{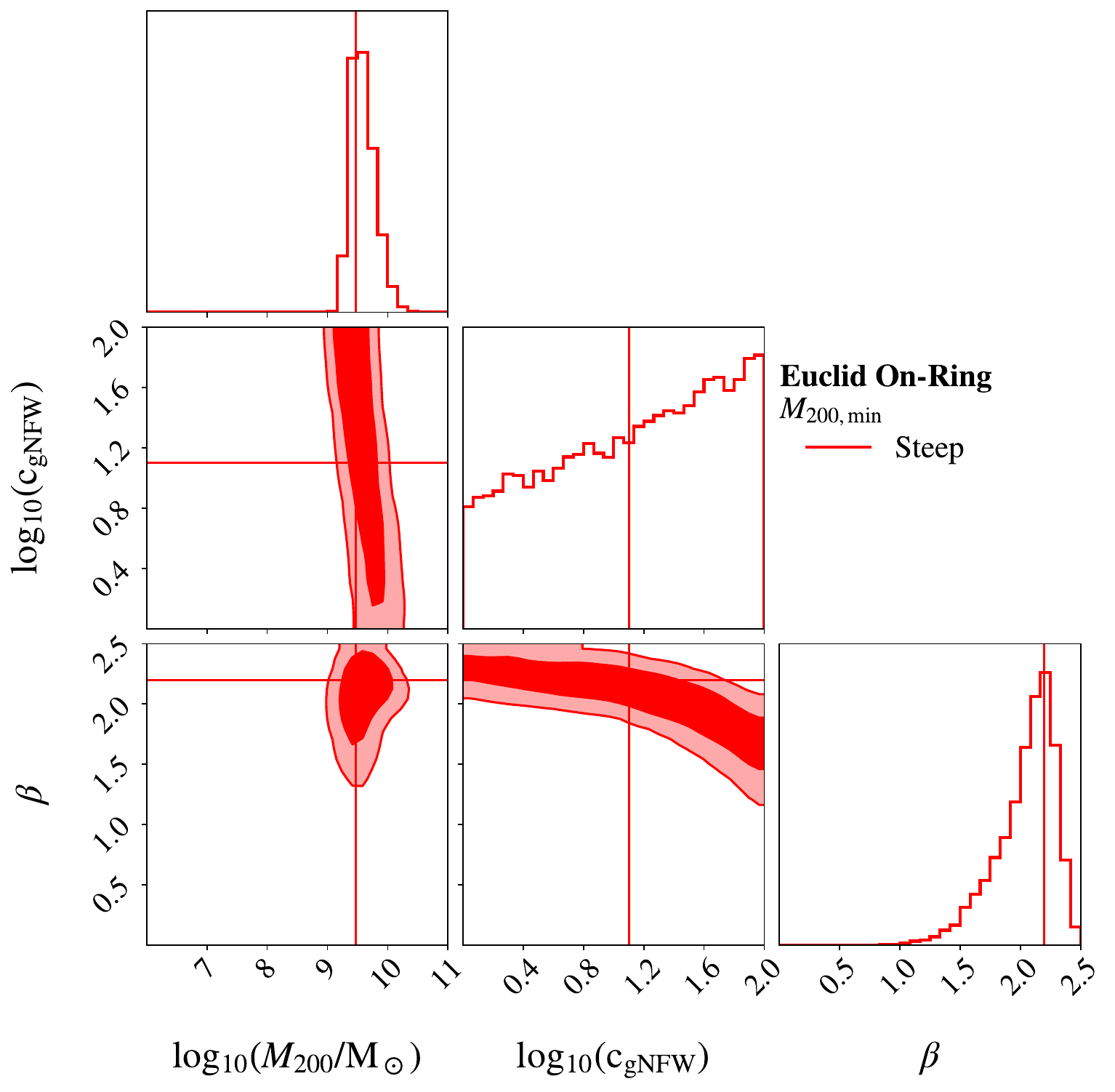}
    \caption{Same as \figref{fig:HST_OnRing_cornerplots}, except for the Euclid On-Ring suite.}
    \label{fig:Euclid_OnRing_cornerplots}
\end{figure*}

\begin{figure*}
    \centering
    \includegraphics[width=0.49\textwidth]{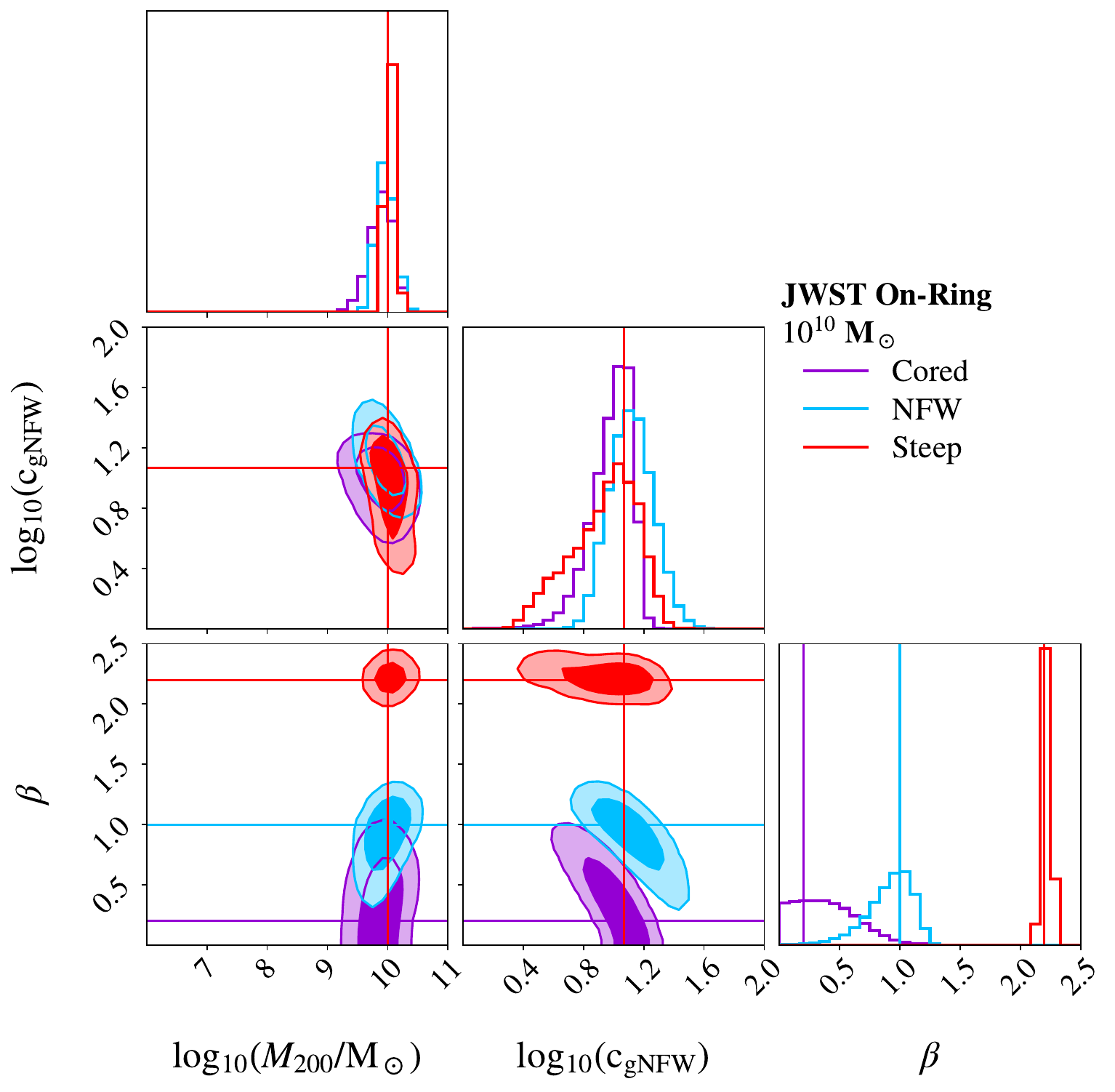}
    \includegraphics[width=0.49\textwidth]{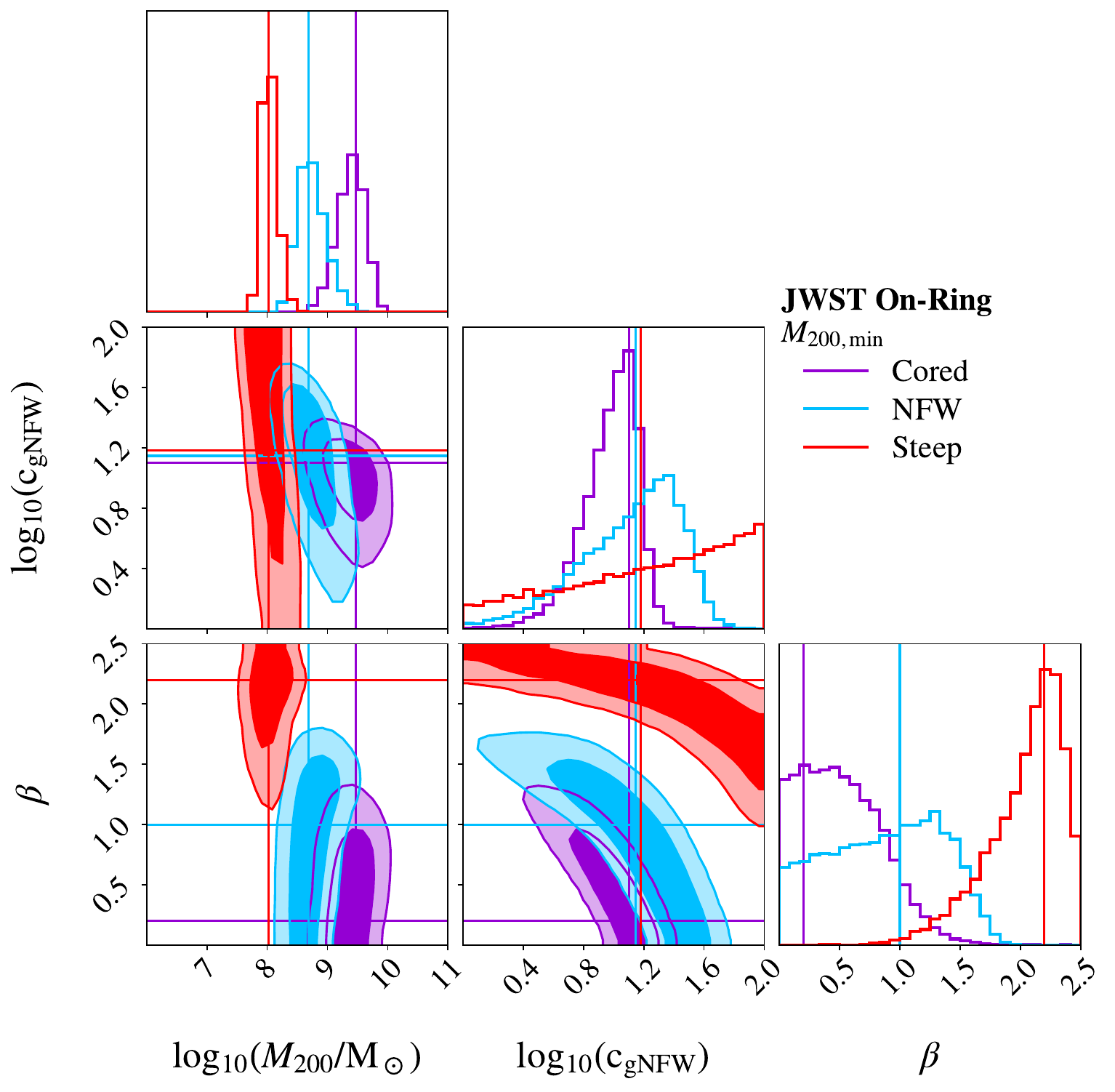}
    \caption{Same as \figref{fig:HST_OnRing_cornerplots}, except for the JWST On-Ring suite.}
    \label{fig:JWST_OnRing_cornerplots}
\end{figure*}

\begin{figure*}
    \centering
    \includegraphics[width=0.49\textwidth]{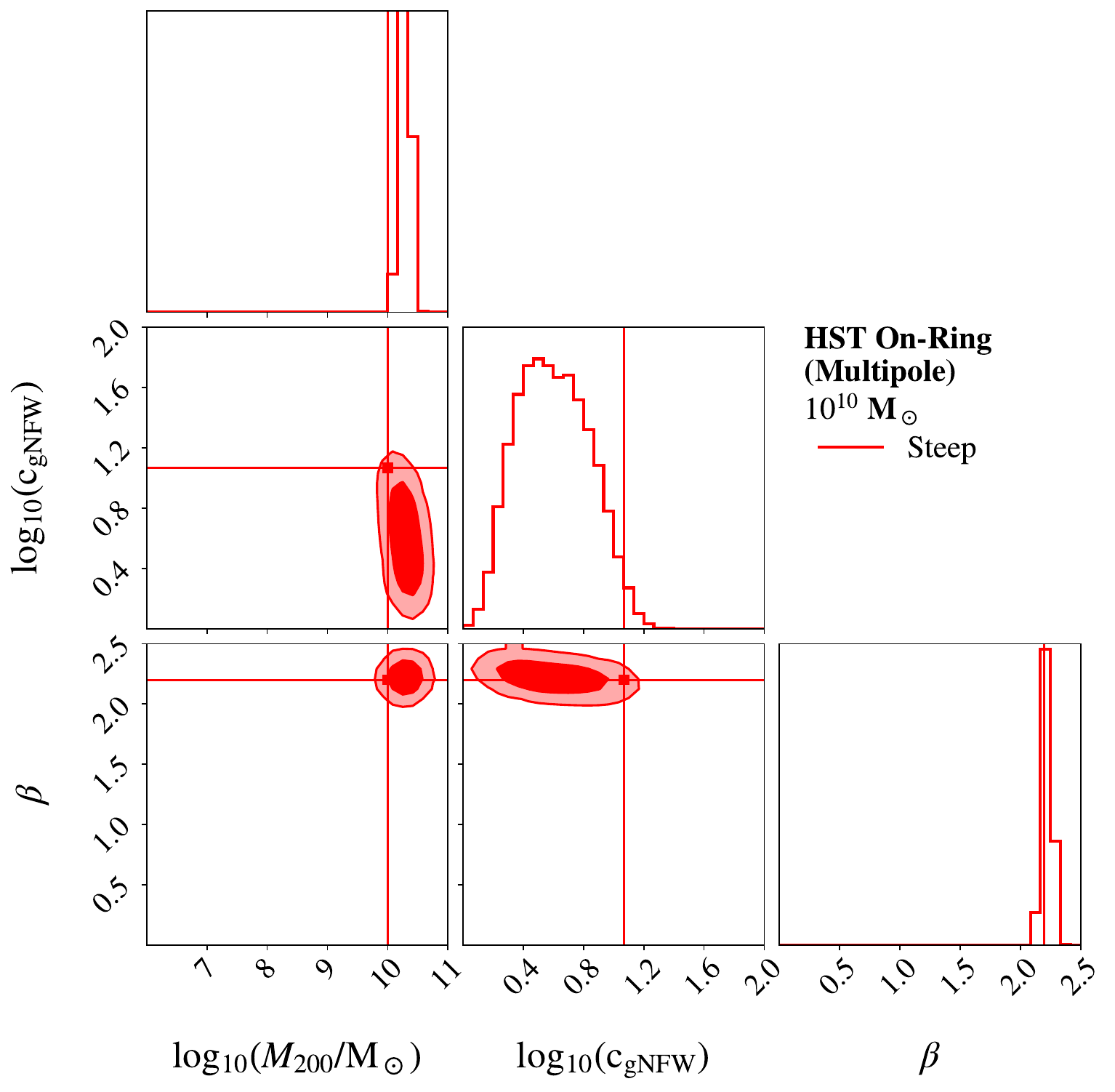}
    \includegraphics[width=0.49\textwidth]{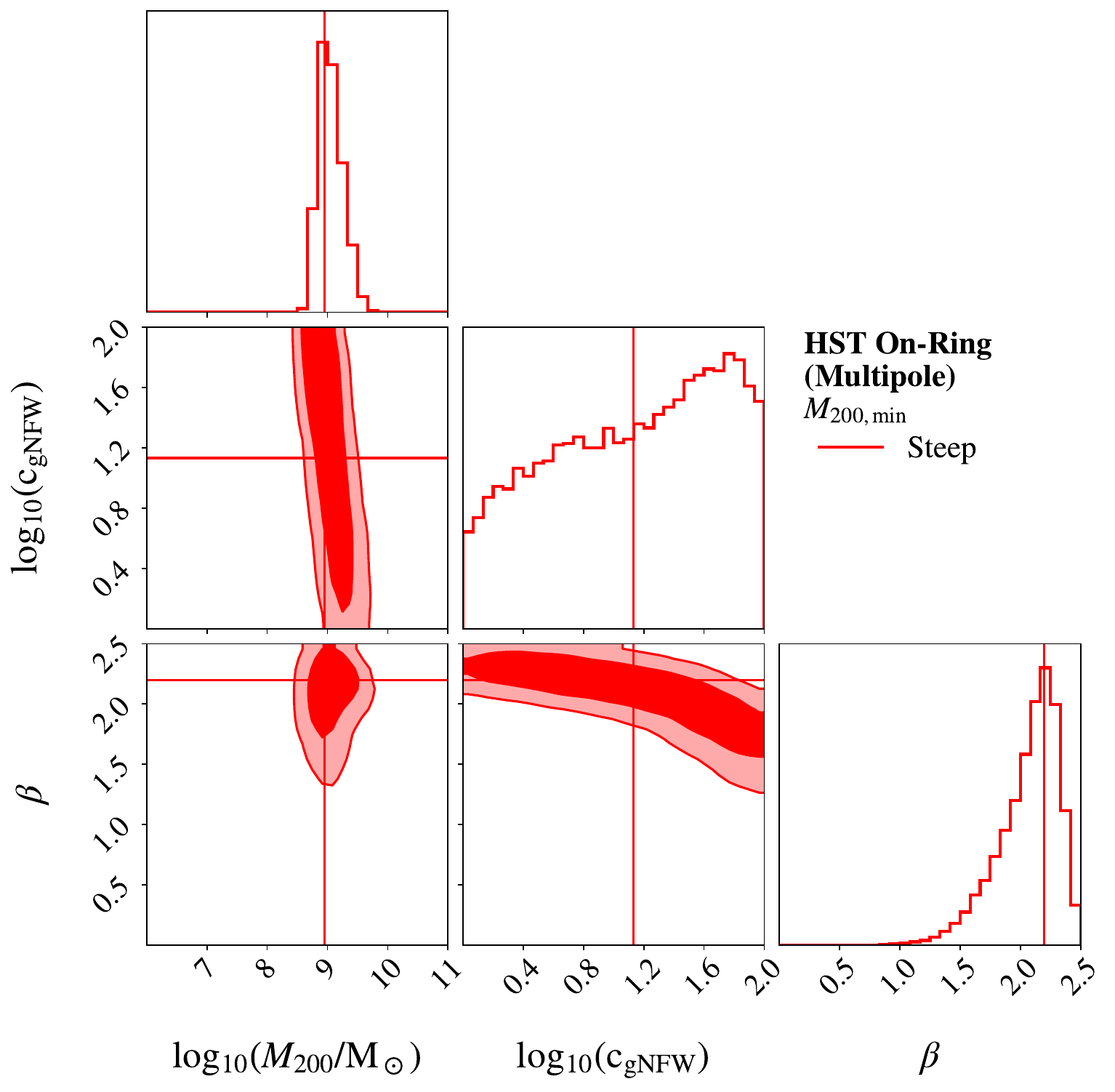}
    \caption{Same as \figref{fig:HST_OnRing_cornerplots}, except for the HST On-Ring~(Multipole) suite. }
    \label{fig:HST_OnRing_m134_cornerplots}
\end{figure*}

This appendix includes supplementary figures that expand on the discussion in the main text. \figref{fig:detectability_plots} compares the detectability for each suite studied in this work against the HST On-Ring results, while  \figref{fig:HST_Offset_cornerplots}--\figref{fig:HST_OnRing_m134_cornerplots} provide the  corresponding corner plots for the $10^{10}$~M$_\odot$ and minimum-mass-detectable subhalos.  The corner plots only show the posteriors for a Cored, NFW, or Steep subhalo if it falls above the detectability threshold. For this reason, the corner plot for the HST Offset~(Multipole) suite is not shown, as all the subhalos are undetectable in that case.

\end{document}